\documentstyle[sprocl,psfig]{article}

\input{psfig.sty}
\def\eps{\epsilon}

\def\be{\begin{equation}}
\def\ee{\end{equation}}
\def\bea{\begin{eqnarray}}
\def\eea{\end{eqnarray}}
\newcommand{\veps}{\varepsilon}  
\newcommand{\nn}{\nonumber}  
\newcommand{\beq}{\begin{equation}}
\newcommand{\eeq}{\end{equation}}
\newcommand{\ket}[1]{| #1 \rangle}                     %  | >
\newcommand{\bra}[1]{\langle #1 \, |}                  %  < |
\begin{document}

%--------------------------------------------------------
%
% list of figures 
%
%  1 fig. 1  magic.eps
%  2 fig  2a levels.eps
%  3 fig. 2b fermi_sphere.eps
%  4 fig. 3a hfbhf1.eps
%  5 fig. 3b hfbhf2.eps
%  6 fig. 4  andreev.eps
%  7 fig. 5  convergence2.eps
%  8 fig. 6a gap_sn.eps
%  9 fig. 6b gap_pb.eps
% 10 fig. 7a rho_sn.eps
% 11 fig. 7b rho_pb.eps
% 12 fig. 8a r2_sn.eps
% 13 fig. 8b r2_pb.eps
% 14 fig. 9a s2n_sn.eps
% 15 fig. 9b s2n_pb.eps
%
%--------------------------------------------------------

\title{ Local Density Approximation for Pairing Correlations in Nuclei }

\author{ \underline{Aurel BULGAC}$^{1,2}$  and Yongle YU$^1$ }

\address{$^1$ Department of Physics,  University of Washington,
Seattle, WA 98195--1560, USA  }

\address{$^2$ Department of Physics, Tohoku University, Sendai 980--8578i, JAPAN}


\maketitle

\abstracts{ We introduce a natural and simple to implement renormalization
scheme of the Hartree--Fock--Bogoliubov (HFB) equations for the case of
zero range pairing interaction. This renormalization scheme proves to be
equivalent to a simple energy cut--off with a position or density dependent running
coupling constant. Subsequently, we present  self--consistent HFB calculations with a
full consideration of the continuum of long chains of spherical tin and lead  nuclei
(essentially from one drip line to another) using the SLy5 interaction. In the pairing
channel we use  a zero range interaction,  treated by means of this new regularization
scheme. }

\section{ Introduction }

\begin{figure}
\psfig{figure=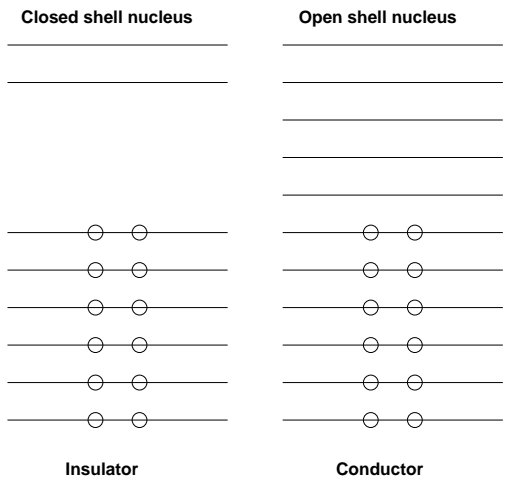,height=2.5in} 
\caption{ Generic single--particle spectra of closed (insulators) and open
                shell (conductors) nuclei. }
\label{fig:mag}
\end{figure}

Since the landmark paper of Bohr, Mottelson and Pines \cite{pines} we have
accumulated so much experimental and theoretical information about pairing in
nuclei that one might be tempted to conclude that ``hardly any rock was left
unturned.'' A closer analysis of the nuclear pairing properties will soon
however reveal how little we really know.  Let us briefly review some basic
facts, which, even though are well known and do not raise any doubt,  will
help  ``drive our point  home.'' In closed shell/magic  nuclei the last
occupied single--particle (sp) state is separated by a gap from the closest
unoccupied sp state, see Fig. (\ref{fig:mag}).  In condensed matter physics 
closed shell nuclei would correspond to insulators, while open shell nuclei to 
conductors. The presence of the gap in magic nuclei is what makes them 
extremely stable. At the same time,  the residual interaction can modify in a 
qualitative manner the structure of the sp spectrum around the fermi level in 
an open shell nucleus. 

%------------------------------------
%   closed and open shell nuclei
%------------------------------------

\begin{figure}
\psfig{figure=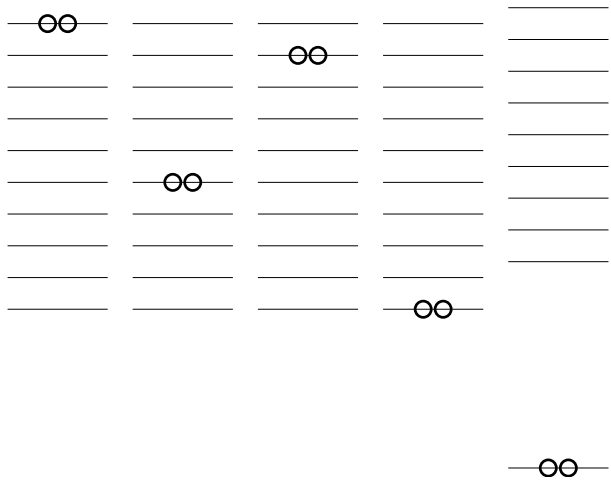,height=2.5in}
\caption{ Cooper's picture of how a Cooper pair if formed. }
\label{fig:cooper}
\end{figure}

The textbook  explanation of  why and how pairing correlations lead to a restructuring of 
the sp spectrum is a simple rehash of the initial Cooper's picture of a Cooper 
pair formation \cite{cooper} and it is illustrated in 
Fig. (\ref{fig:cooper}). The last pair of nucleons, one with spin up, the other
with spin down, can be easily ``pushed around'' by the allegedly weak 
residual interaction, since the spacing between unoccupied levels is
small. The relatively weak residual interaction is effectively enhanced by
the large number of accessible levels and the result is the marked separation
of one of the levels from the rest, see the rightmost level scheme in 
Fig. (\ref{fig:cooper}). 

%------------------------------------
%   Cooper pair
%------------------------------------

Once a gap appears at the fermi surface everybody declares victory,
as in particular  one can then explain why a superconductor has no
resistance, or in other words, why there is no dissipation. A less often asked
question, to which many have  no answer when confronted with it, is: ``How come that a
superconductor conducts at all? Isn't its spectrum similar to the spectrum of
an insulator? A superconductor should be an insulator!? ''

\begin{figure}
\psfig{figure=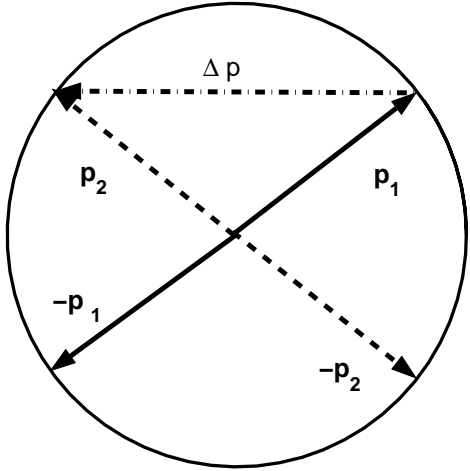,height=2.5in}
\caption{ Scattering of pairs of fermions residing on opposite sides of the
  fermi sphere. }
\label{fig:levels}
\end{figure}

Another picture of the Cooper pair formation is in the momentum
representation, illustrated in  Fig. (\ref{fig:levels}). 
Pairs of nucleons with opposite momenta and spins residing on 
opposite sides of the fermi surface are
``shuffled around'' by a short range interaction all over the fermi
surface. The effect of the interaction is larger the larger the fermi momentum
is, as that corresponds to a larger fermi surface and therefore to
more accessible states. At the same time, scattering particles
from one side of the fermi surface to another requires on average a rather
large transferred momentum, $|{\bf p_1-p_2}| = {\cal{O}}(p_F)$, where
$p_F=\hbar k_F$ is the fermi momentum.  Such a process is possible only if the residual
interaction responsible for this scattering has a sufficiently small radius $
{\cal{O}}(1/k_F)$. Consequently, the pairing interaction is relatively short ranged. 
It is firmly established so far that nuclei belong to the  $s$--wave type of 
``superconductors'' in the weak coupling limit \cite{bcs}, when the paring gap is much 
smaller than the fermi energy, $\Delta\ll \eps_F=\hbar^2k_F^2/2m$.  One can
show that in such a case (in infinite matter) the rms radius of the Cooper pair
\cite{mohit} significantly exceeds the interparticle separation, namely 
$\hbar^2k_F/m\Delta \gg 1/k_F$.  
%------------------------------------
%   Fermi sphere
%------------------------------------

It is easy to see then that under
normal circumstances a Cooper pair has barely any room in an everyday nucleus,
as a Cooper pair  ``wants'' to be larger than a nucleus. This is a remarkable
fact, which for some hard to explain reasons is rarely mentioned in the
nuclear literature and nobody really considered seriously the consequences of
this outstanding feature. As the Cooper pair is a large and a loosely
bound state of two nucleons, its wave function should be described by a single
coupling constant of an essentially zero--range interaction. This is very much
similar to the case of the deuteron, but much closer to the case of two
neutrons (which have an almost bound state), where the scattering
length is  $a=-18.5$ fm.  The theory in such a case should be simple, since
there is a small parameter, the ratio = interaction radius/Cooper pair radius.

The fact that the Cooper pair
is so much larger than the interaction radius should lead to a local pairing
field. For many reasons it is much better to use a local description of the
meanfield properties of many fermion systems. Not only the numerical treatment
is much simpler, but also our intuition works much better when dealing with
local equations. Moreover, there are firm theoretical arguments that a local
treatment is possible and one can devise at least a local Hartree--Fock (HF), or
better, a Kohn--Sham (KS) hamiltonian $h({\bf r})$ \cite{hk,negele,sergei}.  We shall use
the term LDA in the strict KS sense,  as this is the prevailing tendency in the
condensed matter physics and chemistry literature (which by sheer numbers dominate), 
and which slowly becomes an accepted term in this sense in nuclear physics as well. 

The only attempts
to introduce a density functional theory and subsequently a LDA for pairing
known to us are Refs. \cite{oliveira}. However, the LDA extension 
described in these references is in terms
of the anomalous density matrix $\nu ({\bf r}_1,{\bf r}_2)=\bra{gs}\hat{\psi}_\uparrow
({\bf r}_1)\hat{\psi}_\downarrow ({\bf r}_2)\ket{gs}$ and thus in terms of a nonlocal
pairing field as well. If one can adopt the approximation of a zero 
range two--body interaction in the pairing channel as well, then  the HFB 
(or Bogoliubov--de Gennes in condensed matter literature) equations become
%====================================================================
\bea
& &  [h ({\bf r})  - \mu] u_i ({\bf r})
     + \Delta ( {\bf r} )  v_i ({\bf r})
    = E_i u_i ({\bf r} ) , \label{eq:hfb0u}\\
& &  \Delta^* ({\bf r}) u_i ({\bf r})  -
    [ h^* ( {\bf r} ) - \mu ] v_i ({\bf r})
     = E_i v_i ({\bf r}).
\label{eq:hfb0v}
\eea
%====================================================================
Here $u_i ({\bf r})$ and $v_i ({\bf r})$ are the quasi--particle wave
functions,  $\mu$ is the chemical potential,  $\Delta ({\bf r})= -\frac{\delta 
E_{gs}}{\delta \nu^*({\bf r})}$ is the local pairing field, $E_{gs}$ is the ground
state energy  of the system and $ \nu({\bf r})$ is the  anomalous density.
In all the formulas presented here we shall not display the spin
degrees of freedom, but we shall specify the spin degeneracy factor.  

\section{Divergence of the anomalous density for a local pairing field} 

If one takes at face value
Eqs.  (\ref{eq:hfb0u},\ref{eq:hfb0v}) one can show that the diagonal
part of the anomalous density matrix $\nu ({\bf r},{\bf r})$ diverges.
Only the mere fact that the pairing field is local is sufficient enough in order to
arrive at this conclusion. When $|{\bf r}_1-{\bf r}_2|\rightarrow 0$ the 
anomalous density $\nu ({\bf r}_1,{\bf r}_2)$ has the singular behavior
%====================================================================
\beq
\nu ({\bf r}_1,{\bf r}_2)=\sum _i v_i^*({\bf r}_1)u_i({\bf r}_2) \propto
\frac{ 1}{|{\bf r}_1-{\bf r}_2|},
\eeq
%====================================================================
and, as a result,  a local selfconsistent
pairing field $\Delta ({\bf r})$  cannot be defined \cite{ab,note1,bruun}.
One obviously needs a cut--off of some kind, as the divergence  emerges
when the sum extends over the entire HFB spectrum.  
In nuclei and especially in very dilute fermionic atomic systems,
where $k_Fr_0\ll 1$ and $r_0$ is the radius of the interaction, there
is effectively no well defined cut--off and one needs to regularize
the theory.  A finite range interaction will provide a natural cut--off
at single--particle energies of the order of $\veps _c\sim \hbar
^2/mr_0^2$. For such high sp energies the fast spatial oscillations of the quasi--particle
wave functions $u_i({\bf r}), v_i({\bf r})$ will render the nonlocal pairing
field $\Delta ({\bf r}_1,{\bf r}_2)$ ineffective. Even though the
presence of a finite range of the interaction in the pairing channel
formally removes the ultraviolet divergence of the gap, it is very
difficult to come to terms with the fact that a cut--off at an energy
of the order of $\hbar^2/mr_0^2$ could be the responsible for the
definition of the gap both in the case of regular nuclei and very
dilute nuclear matter as well. The characteristic depth of the
nucleon--nucleon interaction potential, which is of the order of
$\hbar^2/mr_0^2$, being the largest energy in the system, can
be effectively considered to be infinite in the case of dilute systems. 
This estimate for the potential energy comes from deuteron properties, 
where the expectation values for the kinetic and potential energies are almost 
equal in magnitude, leading to a small deuteron binding energy. A well
defined theoretical scheme for the calculation of a local pairing field,
should lead to a converged result when only single--particle states
near the Fermi surface are taken into account \cite{best}.

In many treatments of the pairing correlations in infinite systems authors
often underline the dependence of the pairing gap on momentum, that is
$\Delta ({\bf k})$, or in other words, the nonlocality of the pairing
field. We gave already a number of arguments above why a nonlocal pairing, most 
likely, makes little sense.  Let us consider this issue from a slightly 
different point of view. On one hand, typical calculations \cite{elgaroy,khodel} of
the pairing field $\Delta ({\bf k})$ in infinite systems (with no medium polarization
effects taken into account so far)  show that for large momenta
the pairing field decreases, as one would naturally expect.
On the other hand, as soon as the momentum of a quasiparticle state is
sufficiently different from the fermi momentum, when $|k-k_F| \approx
m \Delta (k_F)/\hbar^2 k_F\ll k_F$,
 the effect of the pairing correlations on the
single--particle properties is small, if not negligible \cite{khodel}.  For such momenta,
to a very good
accuracy $E({\bf k})=\sqrt{(\veps({\bf k})-\mu)^2+\Delta^2({\bf k})}
\approx \sqrt{(\veps({\bf k})-\mu)^2+\Delta^2({\bf k}_F)}\approx
|\veps({\bf k})-\mu|$ and thus the use of a $k$--independent pairing field is
a fair  approximation. This is just another way of stating that the
size of the Cooper pair $\hbar^2 k_F/m\Delta$ is much larger
then the average interparticle separation in the weak coupling limit. Typically
this takes place when also the range of the pairing interaction is smaller than
the size of the Cooper pair as well, and thus the pairing interaction could 
and should be described by a single coupling constant.

Most of the calculational schemes suggested so far for infinite
systems reduce to replacing a zero range potential by a
low energy expansion of the vacuum two--body scattering amplitude
\cite{mohit,khodel,blatt,yang,gorkov,randeria,fayans,george,hsu}. The
traditional approach in the calculations of finite nuclei consists
however in introducing a simple energy cut--off, while the pairing
field is computed by the means of a pseudo--zero--range interaction.
In this approach the effective range of the interaction is obviously
determined by the value of the energy cut--off and the two--body
coupling constant in the pairing channel is chosen accordingly\cite{henning}.
Such a pure phenomenological approach lacks a solid theoretical underpinning
and always leaves the reader with a feeling that ``the dirt has been
swept under the rug''.  Another solution, often  favored by other
practitioners is to use a finite range two--body interaction from the
outset, such as Gogny interaction \cite{ring,gogny}.  Besides the fact that
the ensuing HFB equations are much more difficult to solve
numerically, such an approach also lacks the elegance and transparency
of a local treatment and this seemingly simple recipe is indeed as
phenomenological in spirit as the treatment based on a
pseudo--zero--range interaction, with an explicit energy cut--off.
Moreover, in spite of the feeble arguments often put forward in favor
of a finite range interaction in HFB calculations, the only real
argument is the fact that the pairing field would otherwise diverge,
and there is no clear cut mean--field observable which would be noticeable
different in the case of a finite range interaction.
 
It is instructive to show how this divergence emerges and the
simplest system to illustrate this, is an infinite homogeneous one. Since the
divergence is due to high momenta, thus small distances
$|{\bf r}_1-{\bf r}_2|$, this type of divergence is universal and
has the same character in both finite and infinite systems.  Until
recently methods to deal with this divergence were known only for
infinite homogeneous systems
\cite{mohit,khodel,yang,gorkov,randeria,fayans,george,hsu} and only
recently ideas were put forward on how to implement a renormalization
scheme for the case of finite or inhomogeneous systems
\cite{bruun,best,abyy}. 

Assuming for the sake of simplicity that the
spectrum of the HF operator is simply $\veps ({\bf k})=\hbar^2k^2/2m$,
one can represent the anomalous density matrix as follows \cite{ab,bruun,best,abyy}
%------------------------------------------------------------------
\bea
& & \nu({\bf r}_1,{\bf r}_2)=
\int \frac{d^3k}{(2\pi)^3}
\frac{\exp[i{\bf k}\cdot({\bf r}_1-{\bf r}_2)]\Delta }{
2\sqrt{[\veps ({\bf k}) -\mu]^2+\Delta ^2}} \label{eq:n}\\
& & \equiv
\int \frac{d^3k}{(2\pi)^3} \exp[i{\bf k}\cdot({\bf r}_1-{\bf r}_2)]
\left \{
\frac{\Delta }{2\sqrt{[\veps ({\bf k}) -\mu]^2+\Delta ^2}} 
-\frac{\Delta}{2[\veps ({\bf k})-\mu  -i\gamma ]}\right \} \nn\\
& & + \frac{\Delta m \exp ( ik_F|{\bf r}_1-{\bf r}_2|) }{
          4\pi \hbar ^2|{\bf r}_1-{\bf r}_2| },  \label{eq:nu}
\eea
%------------------------------------------------------------------
where $\mu = \hbar ^2k_F^2/2m  $.  The last integral expression is
well defined for all values of the coordinates ${\bf r}_{1,2}$. However, it is
now obvious that the limit ${\bf r}_1\rightarrow {\bf r}_2$ cannot be taken, since
the last term is manifestly divergent. This is the divergence we have to deal
with and  remove it from the theory in some meaningful manner in order to be able to
introduce a local pairing field.  The well known BCS divergence has a different
nature,  is infrared in
character and appears while one approaches the fermi surface, where 
$E({\bf k})=\sqrt{ [ \veps ({\bf k}) -\mu ]^2+\Delta ^2 } $ almost vanishes 
(in the weak coupling limit) and
that leads to an almost logarithmic divergence of the integral in Rel. (\ref{eq:n}).

\section{ Regularization procedure for the anomalous density in finite and
  inhomogeneous infinite systems}

In a nutshell, the regularization of the theory amounts simply to ``throwing
away'' the leading divergent part
$ \Delta m /[4\pi \hbar ^2|{\bf r}_1-{\bf r}_2| ] $ from the rest in the limit
$|{\bf r}_1-{\bf r}_2|\rightarrow 0$. There are several ways to justify this
apparently rather arbitrary procedure: {\it i})  one can use steps
outlined typically in renormalizing the gap equation in infinite
systems -- by relating the divergent part with the scattering amplitude
\cite{mohit,khodel,yang,gorkov,randeria,fayans}-- {\it ii}) or by using
well--known approaches in Quantum Field Theory (QFT) -- either
dimensional regularization \cite{george,hsu}, {\it iii}) or one can
introduce explicit cut--offs and counterterms -- {\it iv}) or
one can follow the pseudopotential approach known for a long time in quantum
mechanics and used among others by Fermi in the 30's.
Irrespective of the approach chosen, one arrives at the
same value for the gap. In particular, if one introduces an explicit cut--off 
one can show that the gap equation becomes
%------------------------------------------------------------------
\beq
 \frac{1}{|g|} =  \int _0^{k_c} dk \frac{k^2}{4\pi^2}
\left [
  \frac{1 }{\sqrt{[\veps ({\bf k}) -\mu]^2+\Delta ^2}}
-\frac{1}{[ \veps ({\bf k}) -\mu - i\gamma ]}
\right ]
  +\frac{ik_Fm }{4\pi \hbar^2},  \label{eq:gapinf}
 \eeq
%------------------------------------------------------------------
where the coupling constant $g$ is defined as
%------------------------------------------------------------------
\beq
g\delta({\bf r}_1-{\bf r}_2)=
\frac{\delta ^2 E_{gs} }{\delta \nu^*({\bf r}_1)\delta \nu({\bf r}_2) }.
\eeq
%------------------------------------------------------------------
and $k_c$ is a momentum cut-off. 
We have assumed here the simplest dependence of the LDA energy density
functional on the anomalous density $\nu({\bf r})$, namely ${\cal{E}}(\tau ({\bf r}) ,
\rho ({\bf r}),|\nu({\bf r})|^2)$, merely for the sake of the
simplicity of the presentation, but more general forms can be used as well. 
In Eq. (\ref{eq:gapinf}) $k_c$ can be
taken to infinity without impunity. In Ref. \cite{best} it is shown that the
minimal value for the cut--off momentum is given by
$E_c={\cal{O}}(\hbar^2k_F^2/2m)$,  where 
$E_c=\hbar^2k_c^2/2m+U-\mu $ and $U$ is the meanfield potential.
Previous approaches
\cite{mohit,khodel,yang,gorkov,randeria,fayans,george,hsu} use
$\veps({\bf k})$ only in the second term instead of $\veps({\bf k})-\mu$
under the integral \cite{note2}. In that case
the last imaginary term does no appear and only then one can relate the
coupling constant $g$ with the scattering length $g=4\pi\hbar^2a/m$ as well.

A note of caution:
it would be incorrect to interpret some of the above formulas in the same manner
as similar looking formulas appearing in various treatments of the pairing
correlations  with a zero--range interaction $V({\bf r}_1-{\bf r}_2)=
g\delta({\bf r}_1-{\bf r}_2)$ (which can be related with the
zero--energy two--particle scattering amplitude $g=4\pi\hbar^2a/m$). As it is
well known for quite some time, even  in the low density region, where
$k_F|a|\ll 1$, there are significant medium polarization corrections
to the pairing gap \cite{gor}. If one adopts a
LDA treatment, then, one is not limited anymore by
similar restrictions on the density. In the
LDA energy density functional the polarization effects are already
implicitly included in the definition of ${\cal{E}}(\tau({\bf r}),
\rho({\bf r}),|\nu({\bf r})|^2)$ and the coupling constant $g$ has no
simple and direct relation to the vacuum two--particle scattering amplitude
$a$. In this sense the LDA is similar in spirit to the Landau fermi liquid theory.

The only attempt to implement a consistent regularization scheme for
finite systems that we are aware of is that of Ref. \cite{bruun}.  In
agreement with the analysis of Ref. \cite{ab} the authors of
Ref. \cite{bruun} conclude that in the case of a zero range two--body
interaction the anomalous density has a $1/|{\bf r}_1-{\bf r}_2|$
singularity. The traditional regularization schemes for infinite homogeneous
systems amounts to subtracting a term proportional to $1/k^2$ in the
gap equation  in momentum representation \cite{mohit}, which in coordinate
representation  corresponds naturally to the same type of divergence  as well
$1/|{\bf r}_1-{\bf r}_2|$. Since the divergence in the anomalous density matrix
$\nu({\bf r}_1,{\bf r}_2)$ is due to large momenta and  thus short distances,
it is not surprising that  the character of the divergence is not affected by the size of
the system. Bruun {\it et al.}
advocate the use of the following calculational procedure for the anomalous
density. First of all one represents the anomalous density as \cite{typo}
%====================================================================
\bea
& & \nu({\bf r}_1,{\bf r}_2) =
\sum _{E_i>0}  \left [ v_i^*({\bf r}_1)u_i({\bf r}_2) +
\frac{\psi _i^*({\bf r}_1)\Delta({\bf r}) \psi _i({\bf r}_2)}{2(\mu -\veps_i)}\right ]  
-\frac{\Delta({\bf r}) G_0({\bf r}_1,{\bf r}_2,\mu)}{2},
\label{eq:bruun} \\
& & [h({\bf r})-\veps_i]\psi _i({\bf r})=0,\quad   
       [\mu-h({\bf r}_1)]G_0({\bf r}_1,{\bf r}_2,\mu)
    =\delta({\bf r}_1-{\bf r}_2) ,
\eea
%====================================================================
where ${\bf r}=({\bf r}_1+{\bf r}_2)/2$.  This particular representation for
$\nu({\bf r}_1,{\bf r}_2)$ was introduced earlier in Ref. \cite{ab}.  One can 
easily justify this subtraction scheme in infinite homogeneous matter, since
$v_i^*({\bf r}_1)u_i({\bf r}_2) =
\Delta \psi _i^*({\bf r}_1) \psi _i({\bf r}_2)/2\sqrt{(\veps_i-\mu)^2+\Delta^2}$.
In the limit ${\bf r}_1\rightarrow{\bf r}_2$ the sum over single--particle states
in Eq. (\ref{eq:bruun}) is converging now and one has only to extract
the regulated part of the propagator $G_0({\bf r}_1,{\bf r}_2,\mu)$,
using the pseudo--potential approach \cite{blatt} 
%====================================================================
\bea
& & \nu_{reg}({\bf r}):=
\sum _{E_i>0} \left [ v_i^*({\bf r})u_i({\bf r}) +
\frac{\Delta({\bf r})\psi _i^*({\bf r})
    \psi _i({\bf r})}{2(\mu-\veps_i)}\right ] \label{eq:nureg1} \\
& & -\frac{\Delta({\bf r})}{2}G_0^{reg}({\bf r},\mu), \label{eq:nureg2}\\
& &G_0^{reg}({\bf r},\mu)= \lim _{{\bf r}_1\rightarrow{\bf r}_2}
G_0({\bf r}_1,{\bf r}_2,\mu) +
\frac{m}{2\pi \hbar^2|{\bf r}_1-{\bf r}_2|}
\label{eq:greg}
\eea
%====================================================================
obtaining for the local pairing field
%====================================================================
\bea
& & \Delta({\bf r})= \frac{4\pi |a|\hbar ^2}{m}
\sum _{E_i>0} \left [ v_i^*({\bf r})u_i({\bf r}) +
\frac{\Delta({\bf r})\psi _i^*({\bf r})
    \psi _i({\bf r})}{2(\mu-\veps_i)} \right ] \label{eq:delreg1} \\
& & - \frac{4\pi| a|\hbar ^2}{m}
\frac{\Delta({\bf r})}{2}G_0^{reg}({\bf r},\mu), \label{eq:delreg2}
\eea
%====================================================================
where $a$ is the two--particle scattering length ($a<0$).
As one can see, the regularization procedure and the extraction of the
regulated part from various diverging quantities is completely
analogous to the familiar procedures in QFT, with the
only difference that in this case everything is performed in coordinate space.
One literally "throws away" the diverging terms and retains the nonvanishing
finite contributions.

\begin{figure}
\psfig{figure=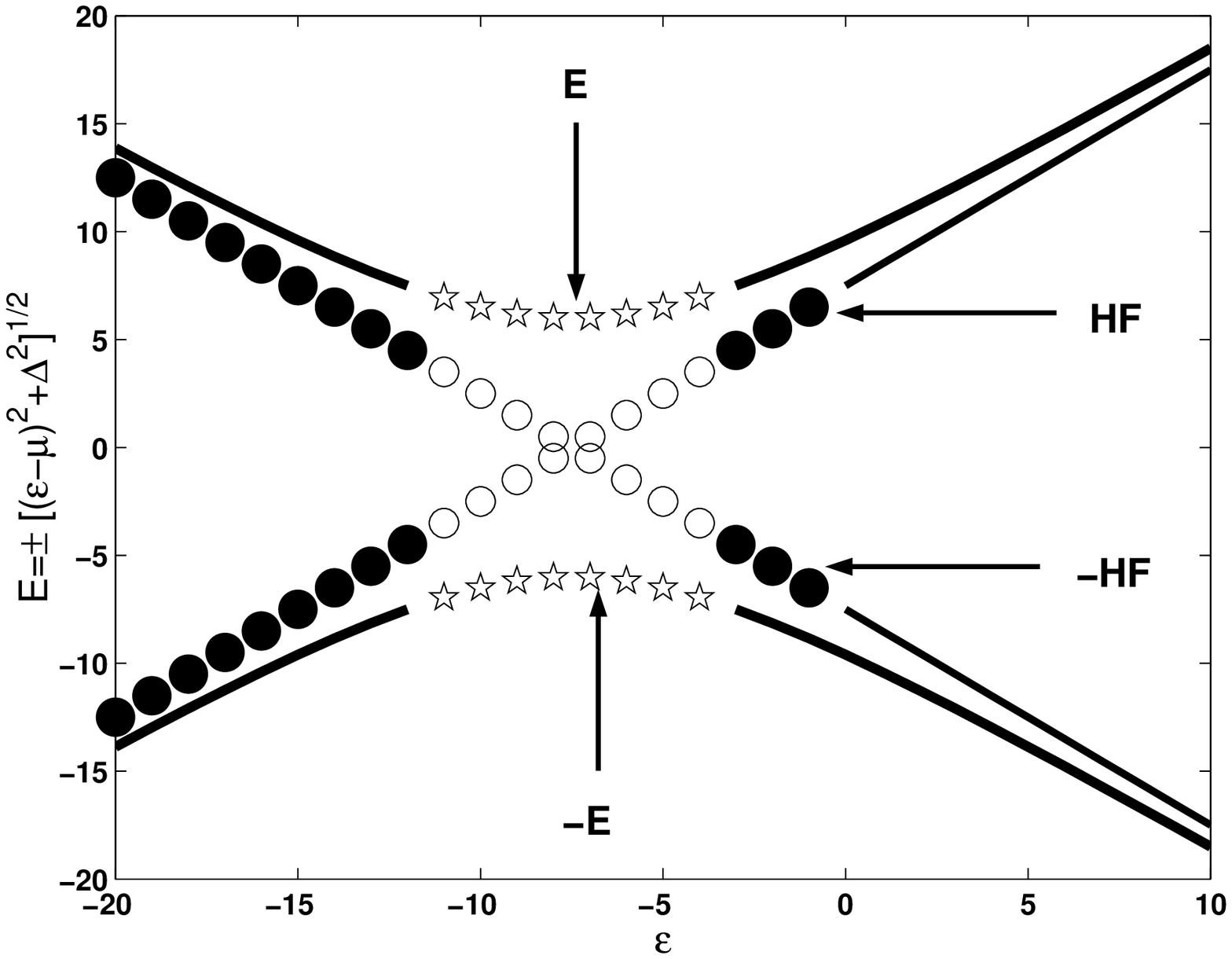,height=2.5in} 
\psfig{figure=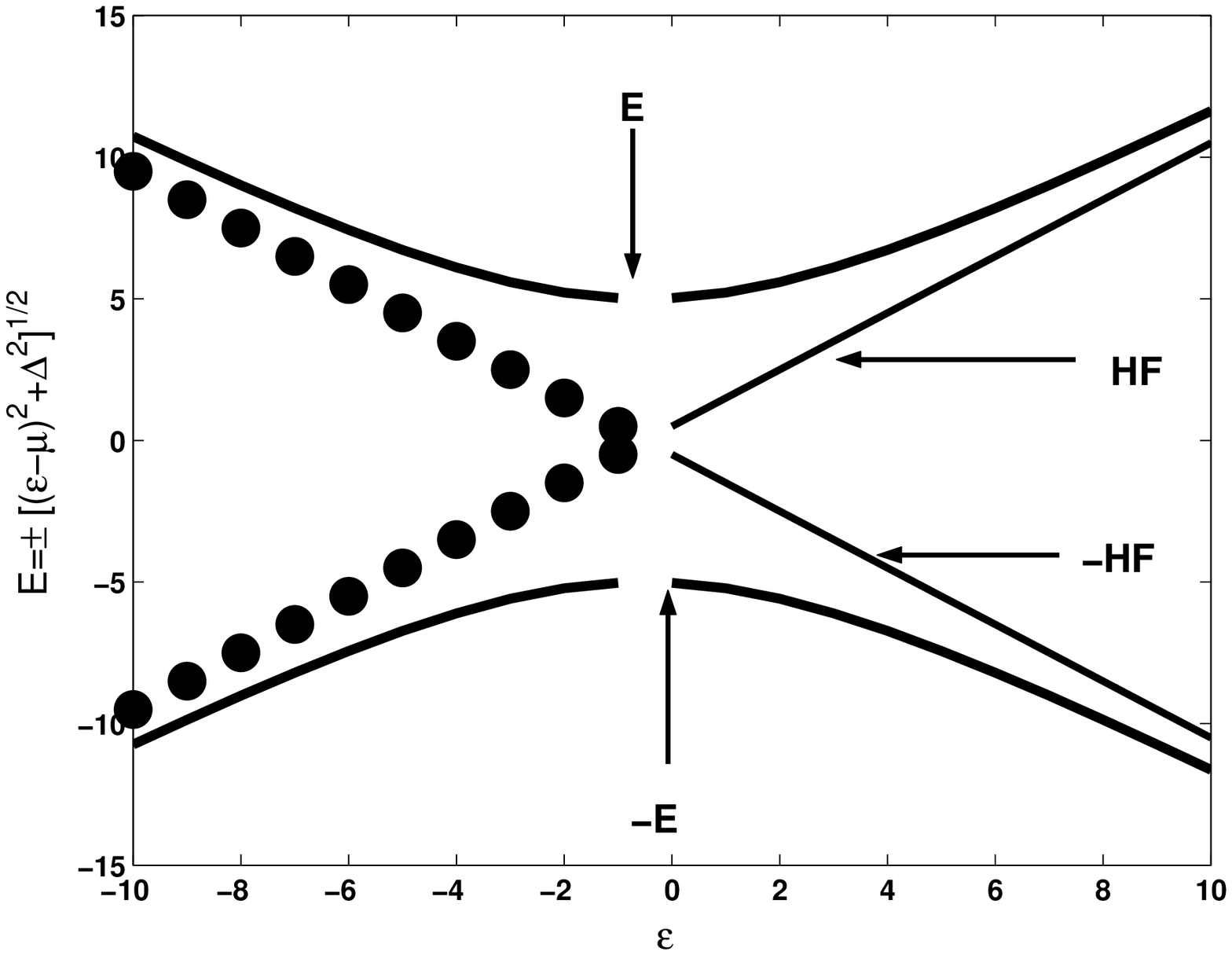,height=2.5in}
\caption{ The empty and filled dots correspond to discrete states of
  Eqs. (\ref{eq:hfb0u},\ref{eq:hfb0v}) for $\Delta\equiv 0$ (marked with HF and
  -HF respectively), and the continuing
  straight lines to the corresponding continuous spectra. 
  The HFB spectra are marked with E and -E, 
  and we use  pentagrams for discrete states and continous
  lines for the respective continous parts of the spectra.
  Only for the discrete HF states marked with empty circles one can find
  corresponding discrete HFB states. }
\label{fig:hfb12}
\end{figure}

There are however problems with using this apparently meaningful regularization
procedure. As it is formulated, the approach of Ref. \cite{bruun} works for
systems in a harmonic trap only and does not apply to atomic nuclei or other
self--sustaining systems. Nobody would argue that pairing correlation depend on
the sp properties in some, hopefully small,  neighborhood of the fermi surface only.
Consequently, after the divergence has been removed, what was left, namely  the
rhs of Eqs. (\ref{eq:nureg1}, \ref{eq:delreg1}), should be defined entirely in terms
of sp properties at and around the fermi level. The problem is that
around the fermi level one cannot establish a 1--1 correspondence between the
HF and HFB properties \cite{ab}.  In order to arrive at these expressions the
authors of Ref. \cite{bruun} used explicitly this 1--1 correspondence in order
to remove in a controlled manner the divergent part of the anomalous density matrix
$\nu ({\bf r}_1,{\bf r}_2)$.  In Fig. (\ref{fig:hfb12}) the nature of the HFB
spectrum around the fermi surface is
illustrated for two cases: the upper panel corresponds to a situation resembling nuclei along
the $\beta$--stability valley ($|\Delta| < |\mu|$)  while  in the lower figure
we show an HFB spectrum for nuclei very close to a
nucleon drip line ($|\Delta| \ge |\mu|$). In the first case one can see that not all discrete HF
levels have a discrete counterpart in the HFB spectrum,  
even in the interval  $2\mu <\veps _i<0$.
In the second case, the HFB equations have no discrete spectrum whatsoever.

%------------------------------------
%   spectra
%------------------------------------

%------------------------------------
%       Andreev 
%------------------------------------
\begin{figure}
\psfig{figure=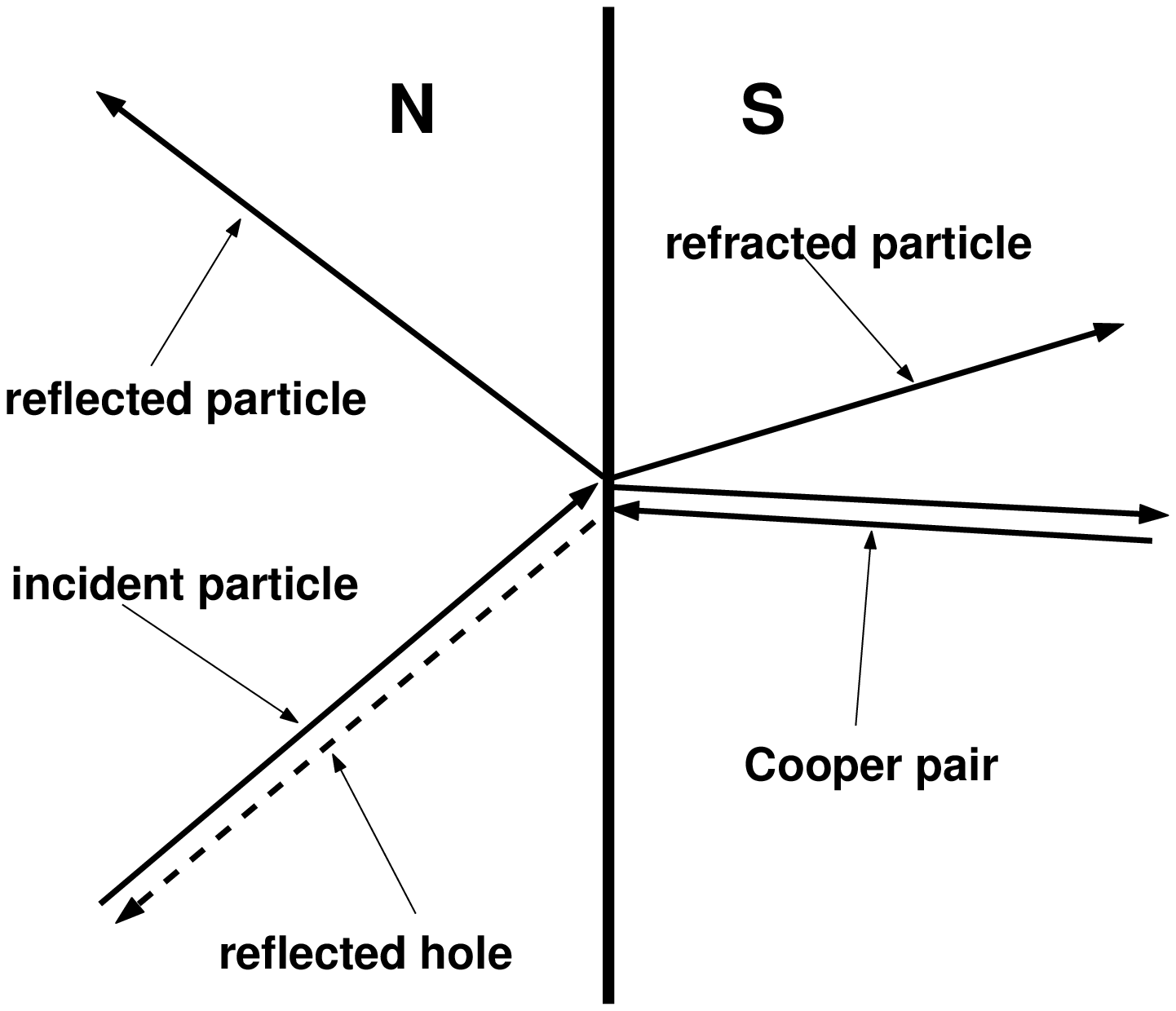,width=3in}
\caption{ Andreev reflection.}
 \label{fig:andreev}
\end{figure}

The fact that hole--like states in HFB have a  the continuous spectrum is known
for quite some time \cite{ab,jacek}.  The explanation given in these references
is  somewhat formal and lacks a simple physical intuition and a pictorial 
representation. As a matter of fact this feature can be
explained in rather simple terms, as the phenomenon has similarities with the so
called Andreev reflection, known for quite some time in condensed matter
physics. Let us imagine that a system is made half of a normal metal and the
other half is a superconductor, see Fig. (\ref{fig:andreev}). In some respects 
nuclei with pairing correlations are like that, as they can be thought of drops of
``superconducting nuclear matter'' embedded in a perfect conductor, the vacuum.  
If a particle coming from the normal part, impinges of the interface, normally
one can expect that with some probability this particle is reflected and with
some other probability the particle is refracted. If however the other side is
superconducting, then there is one more possibility, the particle can get to the
other side, but on the way it picks--up another particle from the normal side and
leaves behind a hole. If the initial impinging particle had the momentum ${\bf
  p}_1$, it will pick another particle with momentum $-{\bf p}_1$ 
and opposite spin and it will form a Cooper pair. Since Cooper pairs are (almost) bosons,
the probability that such an event occurs is bigger, the larger the Cooper pair
condensate is, as in lasers for example. By picking--up an additional particle from
the normal side this process leads to a hole, which is reflected exactly in the
opposite direction of the picked--up particle. What one achieves in this way is a
coupling between particle and hole states. Since the hole states are
coupled with continuum particle states, hole states acquire a continuum character as well.

Now, returning back to the regularization procedure suggested in
Ref. \cite{bruun} one can see that we have a serious problem. In order to
arrive at Eqs. (\ref{eq:nureg1},\ref{eq:nureg2},\ref{eq:delreg1},\ref{eq:delreg2}) 
one had to assume that for each HFB term there was a unique HF term. Such a
correspondence could in principle be established far away from the fermi
surface, where pairing is ineffective.  One can then represent the anomalous
density matrix as follows, by introducing an explicit cut--off
%--------------------------------------------------------------------
\bea
& & \nu({\bf r}_1,{\bf r}_2) = \sum_{E\le E_c} v_E^*({\bf r}_1)u_E({\bf r}_2)
      \label{eq:n1}   \\
  &-& \frac{1}{2}\Delta \left (\frac{ {\bf r}_1+{\bf r}_2 }{2}\right)
\sum_{\veps-\mu<E_c} \frac{\psi^*_\veps ({\bf r}_1)\psi_\veps ({\bf r}_2)}{\veps -\mu}
                                                          \label{eq:n2}\\
         &+& \sum_{E>  E_c}\left [  v_E^*({\bf r}_1)u_E({\bf r}_2) -
         \frac{1}{2}\Delta \left (\frac{{\bf r}_1+{\bf r}_2}{2}\right)
         \frac{\psi^*_\veps ({\bf r}_1)\psi_\veps ({\bf r}_2)}{\veps -\mu}\right
     ]                \label{eq:n3}\\
         &+&  \frac{1}{2}\Delta \left (\frac{{\bf r}_1+{\bf r}_2}{2}\right)
        \sum _\veps  \frac{\psi^*_\veps ({\bf r}_1)\psi_\veps ({\bf r}_2)}{\veps -\mu },
                                     \label{n4}
\eea
%--------------------------------------------------------------------
where for $E>E_c$ there is a 1--1 correspondence $E\leftrightarrow \veps-\mu$,
but not for the the rest of the spectrum.  However, 
one cannot combine (\ref{eq:n1}) with (\ref{eq:n2}) into a single
expression, be that either an integral or sum (depending on the character of
the spectrum) and eventually combine that with (\ref{eq:n3}) to express the entire
quantity as a single sum/integral, independent of $E_c$ as was done in
Ref. \cite{bruun}. Without introducing an
explicit cut--off, beyond which such a 1--1 correspondence could be established,
and treating in some other manner the rest of the spectrum, it is completely
unclear how one should proceed in order 
to calculate Eqs. (\ref{eq:nureg1},\ref{eq:delreg1}), which depend on
the HF and HFB sp properties around the fermi level only.

A related difficulty with the approach suggested in Ref. \cite{bruun} and to a large
extent an even more serious issue  is however the
calculation of  Eqs. (\ref{eq:nureg2},\ref{eq:delreg2}). This  requires the
extraction of the regularized part of the HF propagator for a
potential of arbitrary shape. These two issues are intimately related, as the
final answer is only the sum of Eqs. (\ref{eq:nureg1}) and (\ref{eq:nureg2})
or of Eqs. (\ref{eq:delreg1}) and (\ref{eq:delreg2}) respectively. One has
obviously some freedom here on how to chose the particular form of the subtracted 
quantity, as the regulator is not uniquely defined. Any regulator with the
general structure
%------------------------------------------
\beq
\frac{m\Delta ({\bf r}) F({\bf r}_1,{\bf r}_2)}{2\pi \hbar^2|{\bf r}_1-{\bf r}_2|} ,  
\quad {\mathrm{where}} \quad F({\bf r},{\bf r}) \equiv 1,
\eeq
%------------------------------------------
and otherwise arbitrary function $F({\bf r}_1,{\bf r}_2)$ would be adequate.
The question is: ``How should one choose this regulator?'' The
extraction of the regular part of the propagator is trivial in the case of free motion, and
one more case, the spherical harmonic potential was worked out in Ref. \cite{bruun}.   
We have tried to devise numerical procedures to perform such an
extraction for an arbitrary potential, but the numerical accuracy we could
achieve  was unsatisfactory.  Until such a procedure is suggested, the only
alternative one is left with is to devise a new regulator, which, maybe, could be
handled accurately either numerically or analytically. 

Fortunately a very
simple solution indeed exists. One has to recognize again the physics, namely
that the divergence is ultraviolet in character and thus has nothing to do with
one particular geometrical shape or spatial extent of the 
meanfield potential. This also suggests that the
problem can most likely be handled in a local manner. First of all we
introduce an explicit energy cut--off $E_c$ in evaluating the anomalous
density. In this way we ``divorce'' the HFB and HF sums in Eqs.
(\ref{eq:bruun},\ref{eq:nureg1},\ref{eq:delreg1})  and evaluate them
separately, irrespective of
the existence of the 1--1 correspondence discussed above.
The final result should be  independent of  $E_c$, if this is
chosen appropriately, specifically, if $E_c$ is large enough, then the quantity
given by Rel. (\ref{eq:n3}) is negligible. 
Secondly, we remark that there is no compelling reason to
use the exact HF sp wave functions, energies and propagator in Eqs.
(\ref{eq:bruun},\ref{eq:nureg1},\ref{eq:nureg2},\ref{eq:greg},\ref{eq:delreg1},\ref{eq:delreg2})
and in order to construct the regulator we use a Thomas--Fermi
approximation for the relevant quantities. Since the divergence has an ultraviolet
character, the Thomas--Fermi approximation is particularly well suited.
Thus we arrive at the following relations
%====================================================================
\bea
& & G_0({\bf r}_1,{\bf r}_2,\mu-U({\bf r})) = -
   \frac{m \exp( ik_F({\bf r})|{\bf r}_1-{\bf r}_2|)}{
     2\pi\hbar^2|{\bf r}_1-{\bf r}_2|}    \nn \\
& &=  -\frac{m}{2\pi\hbar^2|{\bf r}_1-{\bf r}_2|}
    -\frac{ik_F({\bf r})m }{2\pi\hbar^2}
    +{\cal{O}}(|{\bf r}_1-{\bf r}_2|), \label{eq:g0reg} \\
& & \nu_{reg}({\bf r}):=  \nu_c({\bf r})
  +\int _0^{k_c({\bf r})} \frac{k^2dk}{4\pi^2} 
   \displaystyle{ \frac{\Delta({\bf r})}{ \mu -
       \displaystyle{ \frac{\hbar^2k^2}{2m}   }-U({\bf r} )+i\gamma }  }
+\frac{ i \Delta({\bf r})k_F({\bf r})m}{4\pi\hbar ^2}  
 \label{eq:nu1reg} \\
& &= \nu_c({\bf r})
   - \frac{ \Delta ({\bf r}) m k_c({\bf r}) }{ 2\pi^2\hbar ^2 }
   \left \{
         1- \frac{ k_F({\bf r}) }{  2 k_c({\bf r}) }
          \ln \frac{ k_c({\bf r})+k_F({\bf r}) }{ k_c({\bf r})-k_F({\bf r})
}
  \right \}    , \label{eq:nu2reg}     \\
& &    \nu_c({\bf r}) =   \sum _{E_i \le E_c} v_i^*({\bf r})u_i({\bf r}),   
  \\
& & h({\bf r})  = -\frac{\hbar^2{\bf \nabla}^2}{2m}+U({\bf r}), \\       
& &     \mu               =  \frac{\hbar^2k_F^2({\bf r})}{2m}+ U({\bf r}),     \\  
& &     E_c              =   \frac{\hbar^2k_c^2({\bf r})}{2m}+U({\bf r}) -\mu , 
\eea
%====================================================================
where the cut--off energy $E_c$ is chosen sufficiently far away from
the fermi level to insure that the rhs of
Eqs. (\ref{eq:nu1reg},\ref{eq:nu2reg}) is independent of $E_c$
and  $\gamma \rightarrow 0+$ at the end.
If the fermi momentum becomes imaginary
(outside nuclei for example) one can easily show that $ \nu_{reg}({\bf r})$
is still real.  It is useful to introduce an effective coupling constant. Then 
the regularized pairing field has the simple form
%====================================================================
\bea
& & \Delta({\bf r})=-g\nu_{reg}({\bf r})
= -g_{\mathit{eff}}({\bf r})\nu_c({\bf r}),
 \label{eq:gapnreg} \\
& & \frac{1}{ g_{\mathit{eff}}({\bf r})}=
\frac{1}{g} 
  -\frac{m k_c({\bf r})}{2\pi^2\hbar ^2}
\left [ 1
  -\frac{k_F({\bf r})}{2 k_c({\bf r})}
\ln \frac{k_c({\bf r})+k_F({\bf r})}{k_c({\bf r})-k_F({\bf r}) }
    \right ] .  \label{eq:geff}
\eea
%====================================================================
The fact that the effective coupling constant depends on position can be
interpreted either as a position dependent running coupling constant or as a
density dependent running coupling constant. 
For a typical nuclear potential which monotonically increases with the radial
coordinate ($dU({\bf r})/dr >0$) one can show that
$ d g_{\mathit{eff}}({\bf r})/dr >0$, thus the effective pairing interaction
is stronger inside than outside nuclear matter (remember $g<0$). 
This is stark contrast with the behavior one would
get using the popular energy cut--off of a $g\delta({\bf r}_1-{\bf r}_2)$
interaction, namely the vacuum renormalization scheme \cite{henning}.
In this case the effective coupling constant is
%====================================================================
\beq
 \frac{1}{g_{vac}({\bf r})}=\frac{1}{g} -\frac{mk_c({\bf r})}{2\pi^2\hbar^2},
\eeq
%====================================================================
from which follows that $d  g_{vac}({\bf r})/dr <0$, if $dU({\bf r})/dr >0$.
If one uses this vacuum renormalization scheme for the case of an infinite
homogeneous system the gap equation is identical to the traditional
regularization scheme \cite{mohit} and the gap equation can be written as \cite{note2}
%====================================================================
\beq
 \frac{1}{|g|}=\int_0^{k_c}dk \frac{k^2}{4\pi^2}
\left [ \frac{1}{\sqrt{[\veps({\bf k})-\mu]^2+\Delta ^2}} -\frac{1}{\veps({\bf k})}
    \right ],
\eeq
%====================================================================
which should be  compared to Eq. (\ref{eq:gapinf}). One can take the limit
$k_c\rightarrow \infty$, as the integral converges, rather slowly however.

%--------------------------------------------------------------------
\begin{figure}
\psfig{figure=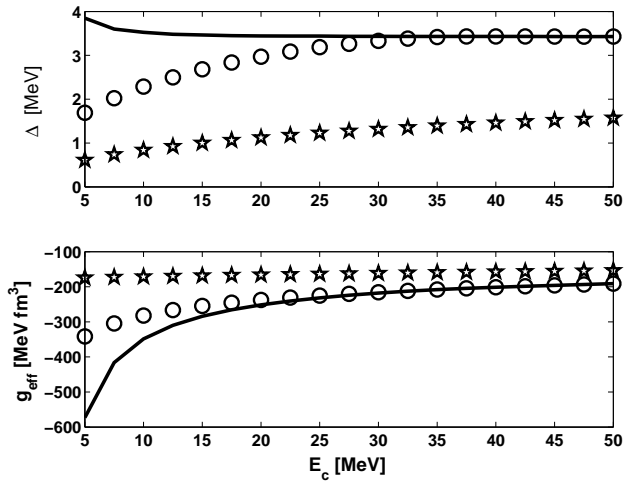,width=3.3in}
\caption{The gap $\Delta$  and the effective coupling
  constant $g_{\mathit{eff}}$ as a function of the  cut-off energy 
  $E_c$ for three regularization schemes. The full lines correspond to
  calculations using the regularization scheme from Ref. $^{12}$, while
  the circles correspond to the regularization scheme presented here.
  The pentagrams correspond to the vacuum regularized coupling 
  constant  $g_{vac} ( {\bf r} )$.  The calculation was performed for
  homogeneous neutron matter with $\rho = 0.08$ fm$ ^{-3} $ and
   $g=-250$ MeV$ \cdot$ fm$^3 $.}
\label{fig:fig2}
\end{figure}
%--------------------------------------------------------------------

In Ref. \cite{best} it was shown that one can introduce an even faster
converging regularization procedure. It is based on the simple observation that
the problematic term in the expression for the anomalous density 
$\Delta/2E({\bf k})$ behaves rather
like $\Delta /2|\veps({\bf k})-\mu|$ instead of   $\Delta
/2\{\veps({\bf k})-\mu\}$.   The main difference
between these two subtraction procedures appears for hole--like states.
Using the ``traditional subtraction scheme''  the
integral over the hole states converges only for
energies of the order of the fermi energy $\eps_F=\hbar
^2k_F^2/2m$. Clearly, in most cases of interest, the so called weak
coupling limit, when $\Delta \ll \eps_F$, there is absolutely no
physical reason to take into account single--particle states so far
away from the fermi level in order to describe nuclear pairing properties.
By simply replacing in the regulator
$\veps({\bf k})-\mu$ with $|\veps({\bf k})-\mu|$ one obtains an equally
simple,  but impressively much faster convergent regularization procedure, see Fig. 
({\ref{fig:fig2}).  Now indeed one can claim that there is an explicit procedure,
which defines the paring field and the anomalous density in terms of sp
properties in the immediate vicinity of the fermi surface only.

A last point, when evaluating the total energy of the system one has to be careful and
calculate the expression \cite{george}
%--------------------------------------------------------------------
\beq
E_{gs}=  \int d^3r \left [
         \frac{\hbar^2}{2m} \tau_c({\bf r})
         -\Delta ({\bf r})\nu_c({\bf r})
                                 \right ] + E_{pot},
\eeq
%--------------------------------------------------------------------
where $E_{pot}$ is the usual HF potential energy contribution, since the
kinetic energy density $\tau_c({\bf r})=2\sum_{E\le E_c} |{\bf \nabla} v_E({\bf
  r})|^2$ diverges in a similar fashion as  $\nu_c({\bf r})$ with $E_c$, but $E_{gs}$ does not.

\section{HFB calculations of spherical tin and lead nuclei with SLy5
  interaction and zero range pairing interaction } 

With a clear regularization scheme in place one can now aim for a 
selfconsistent description of nuclear properties within the HFB method.
Perhaps the most appealing way to proceed is to use the Kohn--Sham LDA
approximation. The philosophy of this approach is to first compute the
properties of infinite homogeneous matter in an {\it ab initio} framework \cite{sergei}, 
in order to derive the energy density functional.  Lacking that, one can resort
to phenomenological approaches. Since we are interested in treating open shell nuclei, this
functional should depend on the kinetic energy  
$\tau({\bf r})=2\sum _i |{\bf \nabla} v_i({\bf r})|^2$
the normal $\rho({\bf r})=2\sum _i |v_i({\bf r})|^2$
and the anomalous $\nu ({\bf r})=\sum _i v_i^*({\bf r})u_i({\bf r})$
densities. (The angular momentum density ${\bf J}({\bf r})$ is naturally
included as well in calculations, along with other necessary ingredients such
as Coulomb interaction.) 
For the sake of simplicity of this initial exploratory survey of nuclear
properties using this new regularization scheme, we shall choose a rather simple
parametrization of the energy density functional as follows
%--------------------------------------------------------------
\beq
{\cal{E}}_{LDA}({\bf r})={\cal{E}}_0({\bf r})+g|\nu ({\bf r})|^2,
\label{eq:enden}
\eeq
%--------------------------------------------------------------
where for energy density describing normal systems we have chosen
${\cal{E}}_0({\bf r})$ given by the so called SLy5 interaction
\cite{sly}. We could have also used the parametrization suggested by Fayans
\cite{sergei}, constructed directly from properties of infinite homogeneous symmetric
nuclear matter and pure neutron matter \cite{fpwff}. A similar constraint was
imposed for the SLy interaction. As for the dependence on the
anomalous density no such information is available yet. There are quite a number
of calculations of pairing properties of homogeneous  neutron and symmetric
nuclear matter, see Refs.  \cite{elgaroy,khodel,garrido} and further references
therein.  The usefulness of all these results is
questionable however at this time. 
Some authors would try to convince the reader that
the use of a bare NN--interaction in the pairing channel is appropriate
\cite{elgaroy}. Others would not quite subscribe to this point of view, but in
the absence of any meaningful theoretical input would proceed with pairing
calculations using bare NN--interaction anyway \cite{khodel,garrido}.  Even
though there is a clear danger of double--counting, in some instances the Gogny
interaction is used as well \cite{garrido}, claiming, however, that in this
particular channel the Gogny interaction is perhaps closer to a bare
NN--interaction rather then to a G--matrix, as was initially envisioned by its
creators \cite{ring,gogny}.  The list of uncertainties however does not stop
here. Assuming for the moment that indeed one can use the bare NN--interaction
in the pairing channel, this is not going to help us too much. One might naively
suspect that in very dilute matter medium polarization effects are negligible
and thus one can indeed use the vacuum scattering amplitude to evaluate the
pairing gap in the leading order in $k_F|a|$ expansion. That would correspond
formally to the energy density functional we have chosen above, see
Eq. (\ref{eq:enden}). The medium polarization effects are however not
negligible and, moreover, depending on the composition of the system they can
either enhance significantly the pairing correlations in symmetric nuclear
matter by a factor of $(4e)^{1/3}\approx 2.2$
or suppress them by the opposite factor $(4e)^{-1/3}\approx 0.45$
in the case of pure neutron matter.  What happens at finite densities is
so far a big unknown, as no comprehensive study of the isospin dependence of
the pairing properties in infinite matter was performed, as far as we know.

%------------------------------------
%       Delta 
%------------------------------------
\begin{figure}
\psfig{figure=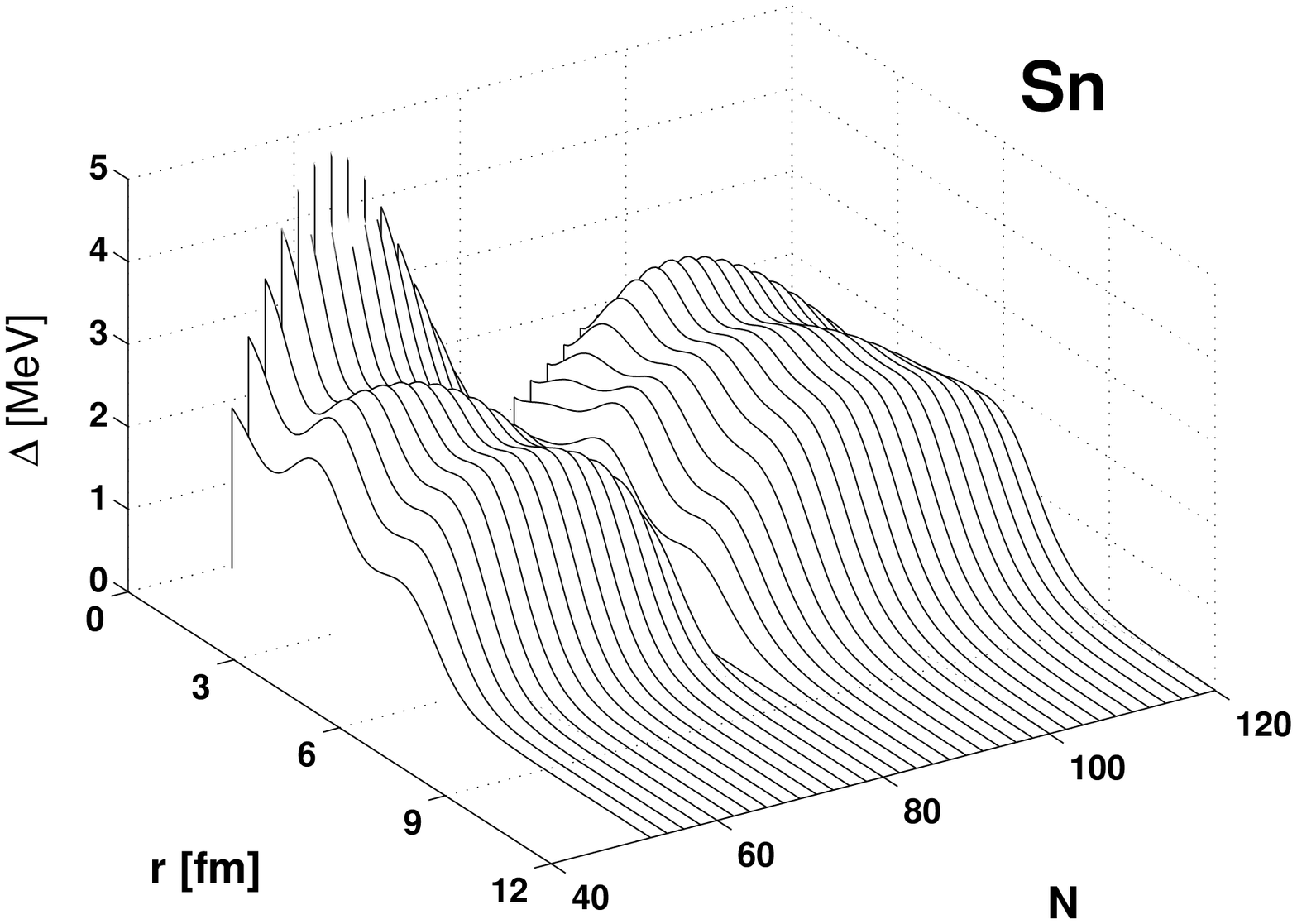,width=3.5in}
\psfig{figure=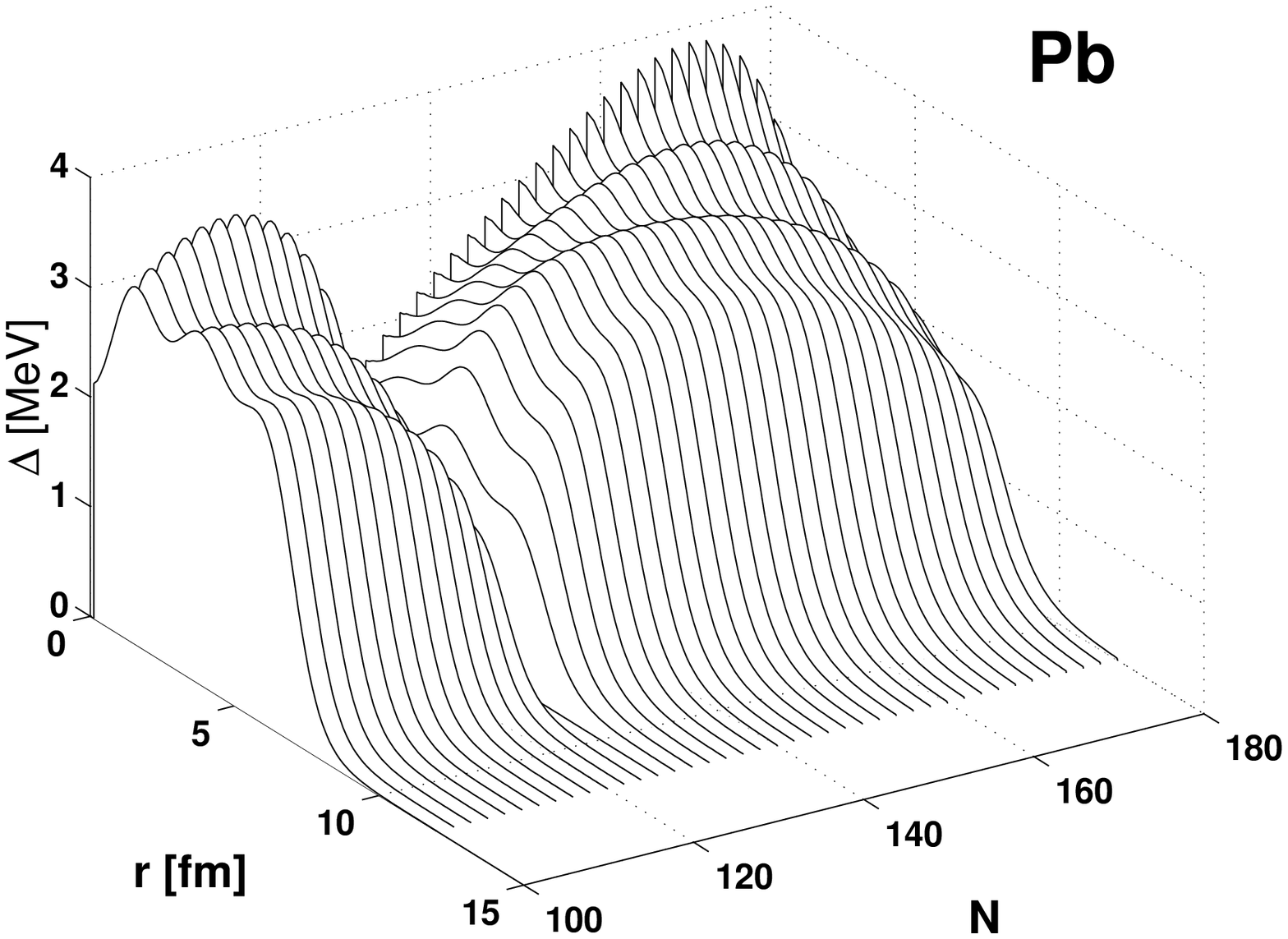,width=3.5in}
\caption{ The pairing field $\Delta({\bf r})$ in spherical even--even tin and lead isotopes. }
\label{fig:gap}
\end{figure}

In principle, if full microscopic calculations of the pairing properties of
infinite homogeneous matter would be available and one would know for example
the dependence of the pairing gap at the fermi level as a function of the
chemical potential, then one could use Eq. (\ref{eq:gapinf}) to extract the density
dependence of the effective bare coupling constant $g$ on density. Ideally we
would like to know both the isoscalar and the isovector nature of the pairing
properties.  In the absence of such input we have resorted to the only avenue
left to us, assume the simplest form for the dependence of the energy density
functional on the anomalous density and try to determine the single constant
$g$ from available experimental information. The authors of Refs. \cite{garrido}
suggested a somewhat similar approach, using as input however results from
infinite matter calculations with bare NN--interaction. Besides the
uncertainties mentioned above with such type of approaches, the
zero--range interaction suggested by these authors is determined by three
parameters, which have a rather unclear and doubtful theoretical
underpinning. One of the reasons is of course the fact that until recently a
meaningful regularization scheme for zero--range pairing interaction  in finite
systems was not available.  We have shown here that such an effective zero--range
interaction can be derived easily and it is fully characterized by a single
effective position/density running coupling constant $g_{\mathit{eff}}$.
A $g_{\mathit{eff}}$ extracted in this way will show an additional density dependence
(besides the one originating from the regularization procedure), arising from the density
dependence of the pairing gap as a function of the density in infinite nuclear matter.

Since we shall consider here nuclei with a magic proton number, pairing will
appear only in the neutron system and thus we can limit ourselves to a single
bare coupling constant $g$. The first part of the nuclear density
functional ${\cal{E}}_0$, apart from the Coulomb interaction, is
isospin invariant, see discussion in Refs. \cite{sergei,vasya}. 
The part describing the pairing correlation is not explicitly an isospin conserving functional
of the anomalous density.  An isospin conserving energy density functional
should have the property (apart from the explicit Coulomb part) \cite{sergei,vasya}
%-----------------------------------
\beq
{\cal{E}}(\rho _n,\rho_p, \nu_n,\nu_p) \equiv {\cal{E}}(\rho _p,\rho_n, \nu_p,\nu_n).
\eeq
%-----------------------------------
We have displayed here only the normal and anomalous density dependence.
Perhaps the simplest parametrization of the pairing part of the energy density
functional  would be  of the form
%-----------------------------------
\beq
{\cal{E}}_{LDA}(\rho _n,\rho_p, \nu_n,\nu_p) =
 {\cal{E}}_0(\rho _n,\rho_p) +g_0|\nu_n+\nu_p|^2 +g_1|\nu_n-\nu_p|^2,
\eeq
%-----------------------------------
where ${\cal{E}}_0(\rho _n,\rho_p) $ describes the normal nuclear properties and
$g_0$ and $g_1$ would  in principle depend on $\rho _n$ and $\rho_p$ as well. We want to
stress here that the energy density functional should be isospin conserving,
which is due to the strong isospin conserving character of the nuclear
forces. However, in various particular ground states of various nuclear systems
the isospin symmetry might and is as a rule broken. This is in complete analogy
with magnetism. The hamiltonian (effective or exact) describing a magnetic system
is always rotational invariant, even though magnetization can spontaneously 
appear in particular systems. There is no {\it a priori} reason however to
conclude that the energy density functional has such a simple dependence on the
anomalous density and we might uncover phenomena which would require the
introduction of at least quartic terms in the anomalous density. So far, it seems
that nuclear pairing fields are relatively weak and quartic terms are not needed.

Limited by these uncertainties, concerning the description of
pairing correlations, we performed a first exploratory calculation of tin and
lead spherical nuclei, essentially from one drip line to the other. We have
solved numerically the ensuing HFB equations in coordinate representations and in
order to treat correctly the continuum spectra of the HFB equations we have
used the Green function technique of Fayans {\it et al} \cite{sergei,fayans} in order to
evaluate the normal and anomalous densities, by integrating  the corresponding 
normal and anomalous sp propagators in the complex energy plane. 
Typical results for the neutron selfconsistent pairing field in these
nuclei are shown in Fig. (\ref{fig:gap}). In Figs. (\ref{fig:rho}) and
(\ref{fig:r2}) we show the neutron and proton density distribution for some
selected nuclei and the corresponding expectation values for $r^2$.
In the results shown in these figures for the tin isotopes we have used a
bare coupling constant $g=-270$ MeV$\cdot$ fm$^3$ and 
$g=-300$ MeV$\cdot$ fm$^3$ for lead isotopes

For those nuclei for which the binding energies have been measured ,
we have compared the two--neutron separation energies
calculated with a range of values for the bare coupling constant $g$ and
compared with experimental data. By varying $g$ the densities are only slightly
affected, while the pairing fields change roughly proportional to the value of
$g$  used. The $S_{2n}$ energies however seem to be reproduced better for 
$g=-300$ MeV$\cdot$ fm$^3$. One has to take these apparently good results
with a grain of salt, as a final fit of the energy density functional should
involve both its normal part (especially its isovector dependence) along with
its pairing properties.  One can however conclude from here  that it is
very likely that one can achieve a very good description of masses, perhaps rms
radii as well, with a single pairing coupling constant, maybe two, to account
for the isospin dependence. More work is required to determine if
any genuine density dependence (not that arising from regularization) is needed
to describe nuclear properties. There is an ongoing debate on whether pairing
has a volume or surface character in nuclei \cite{fayans,jd}, but no clear cut
answer so far.

%------------------------------------
%       rho
%------------------------------------
\begin{figure}
\psfig{figure=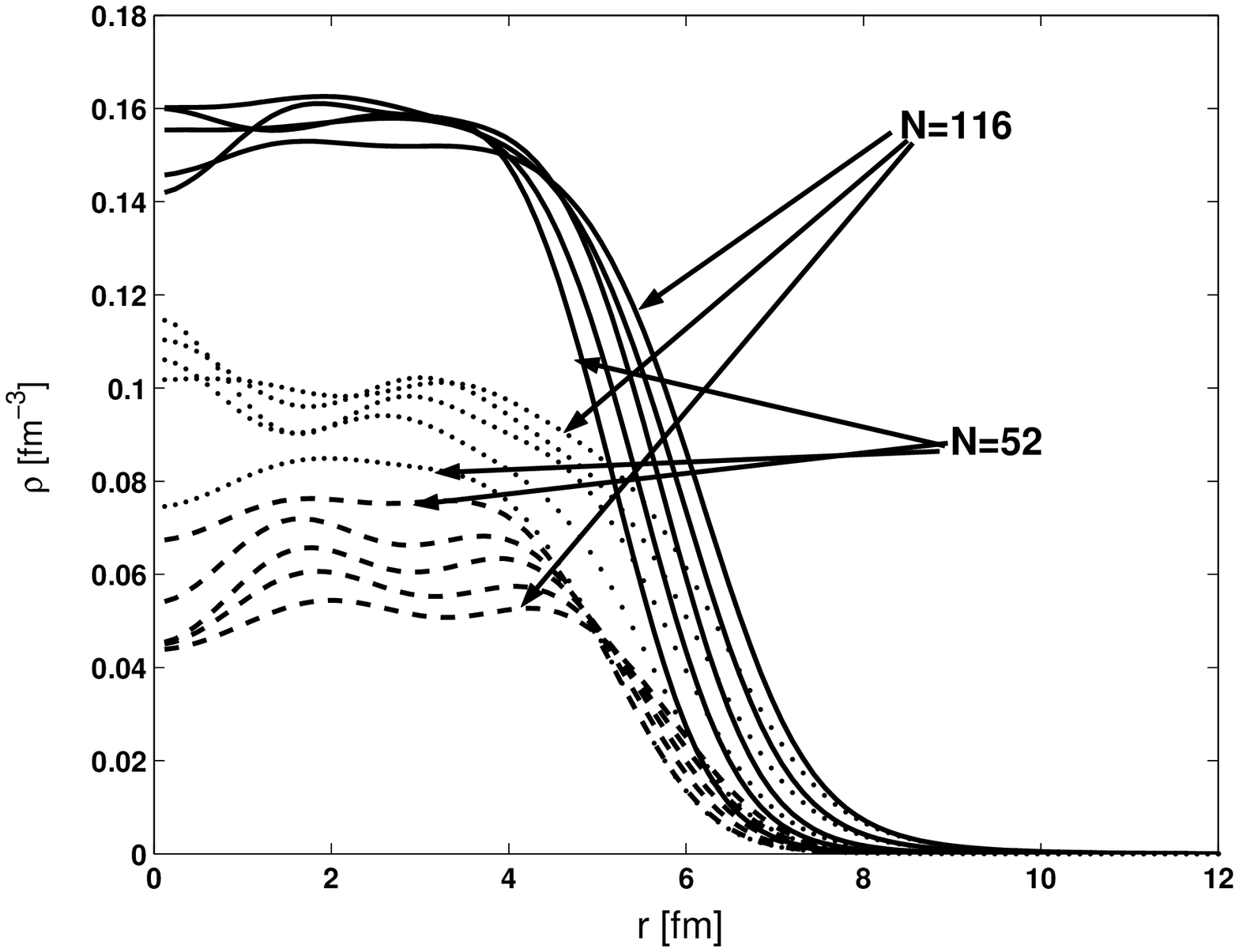,width=4in}
\psfig{figure=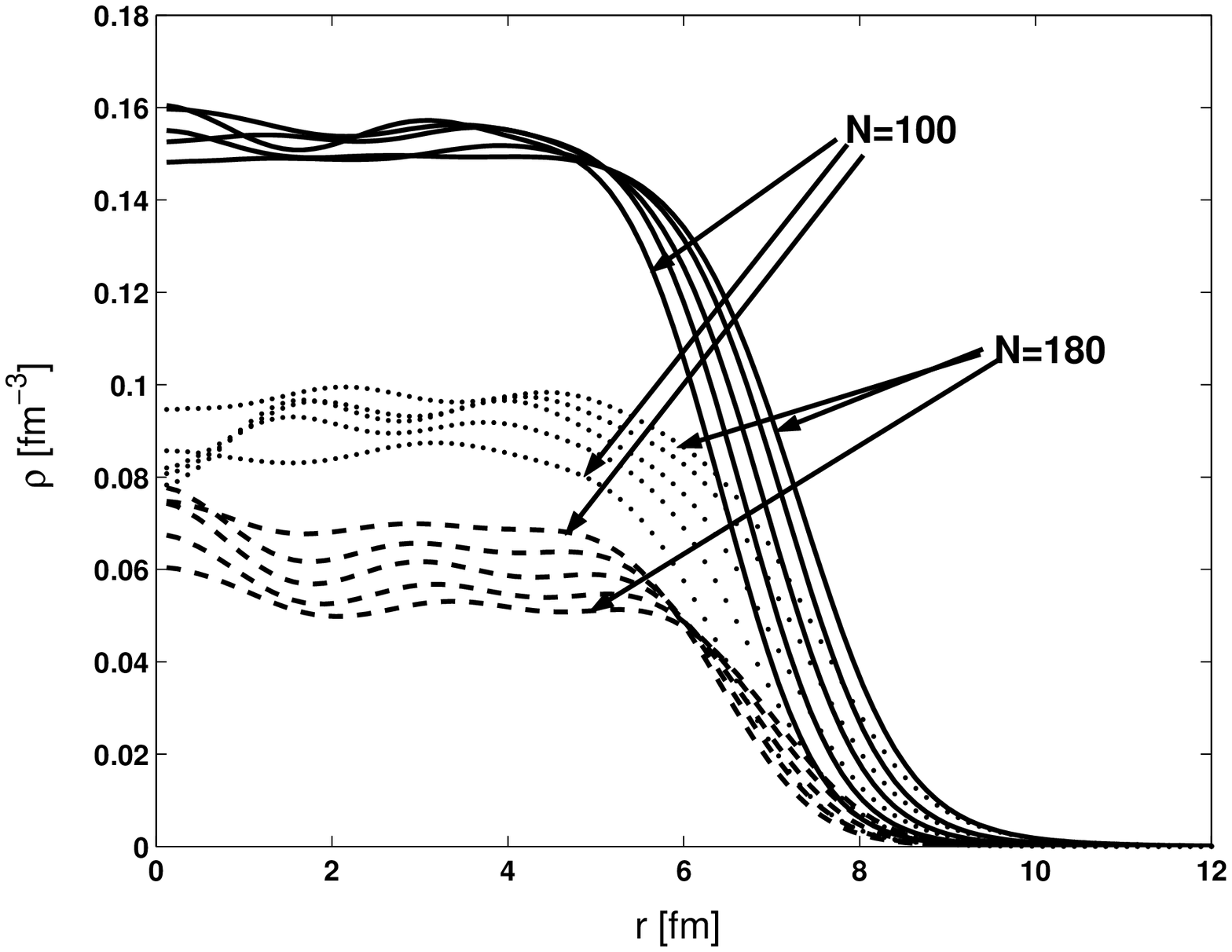,width=4in}
\caption{ The proton (dashes), neutron (dots) and total (full line) densities
  $\rho ({\bf r})$ in tin isotopes for $N=52, 68, 84, 100, 116$ and in lead
  isotopes for $N=100, 120, 140, 160, 180$. }
 \label{fig:rho}
\end{figure}

%------------------------------------
%       r2
%------------------------------------
\begin{figure} 
\psfig{figure=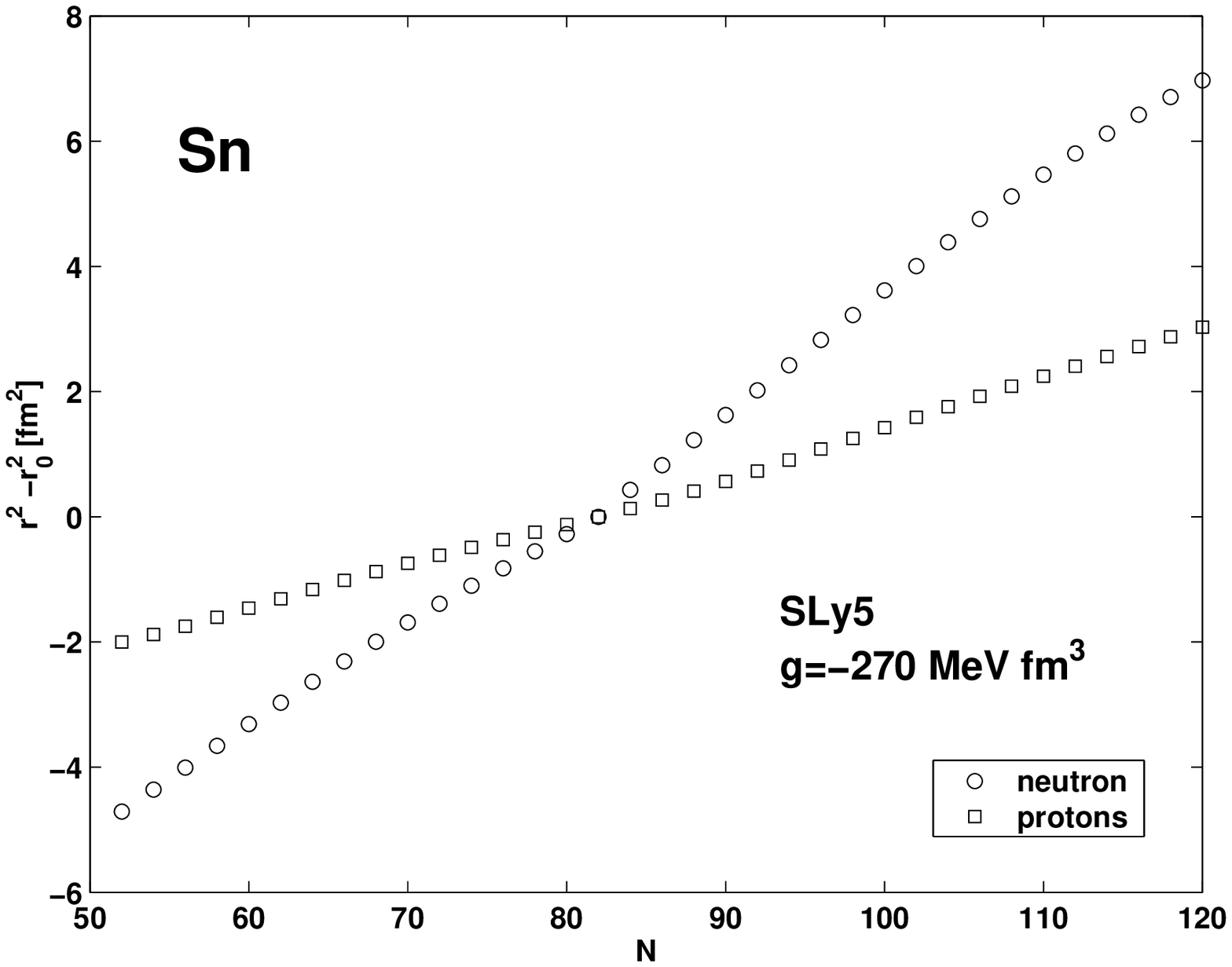,width=4in}
\psfig{figure=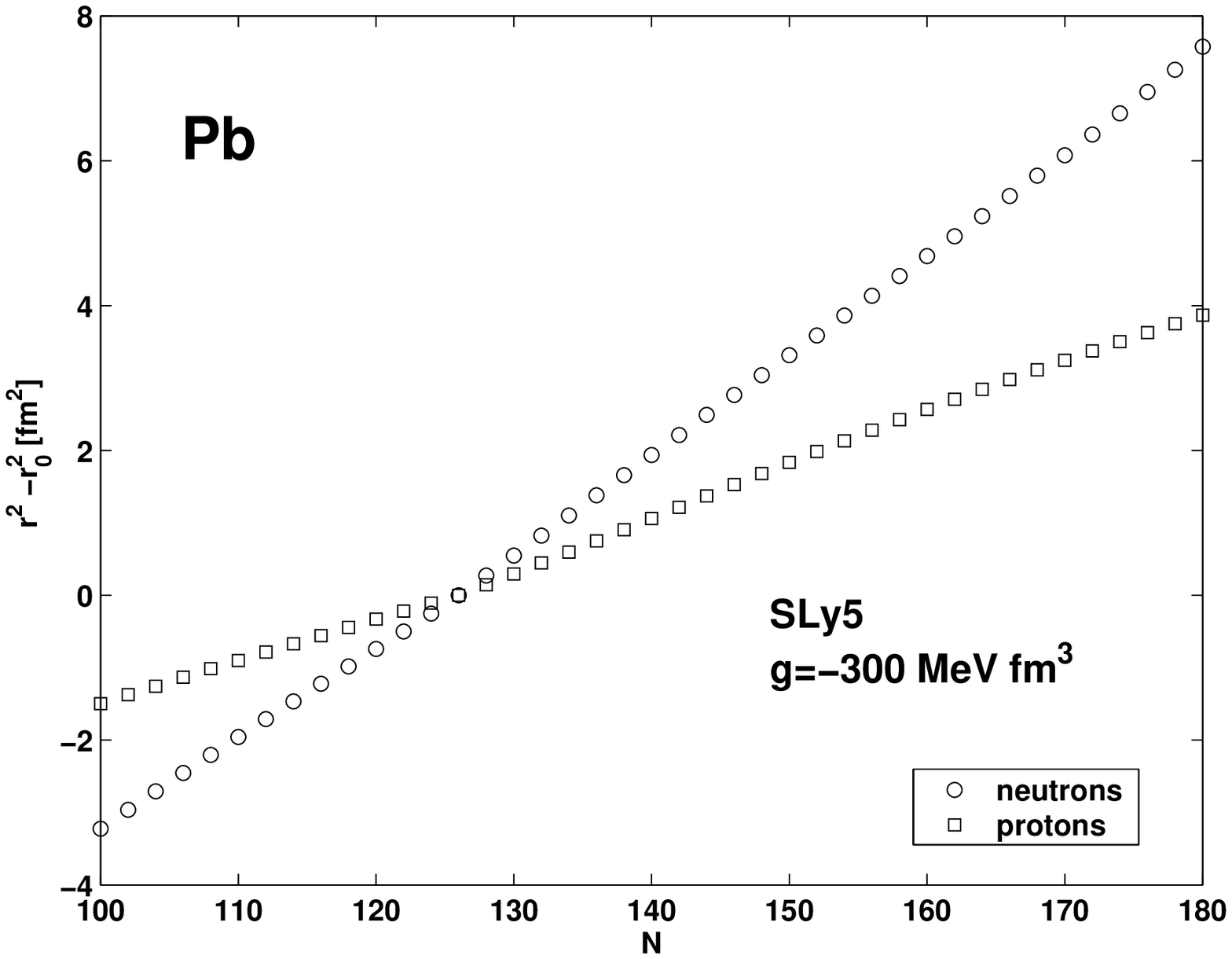,width=4in}
\caption{ The proton and neutron expectation values for  $r^2-r^2_0$ in tin and lead
  isotopes, where for tin isotopes $r^2_0$ were the corresponding values for
  $^{132}Sn$, while for lead isotopes  the values of $r^2$ corresponding to $^{208}Pb$. }
\label{fig:r2}
\end{figure}

%------------------------------------
%       S2n
%------------------------------------
\begin{figure}
\psfig{figure=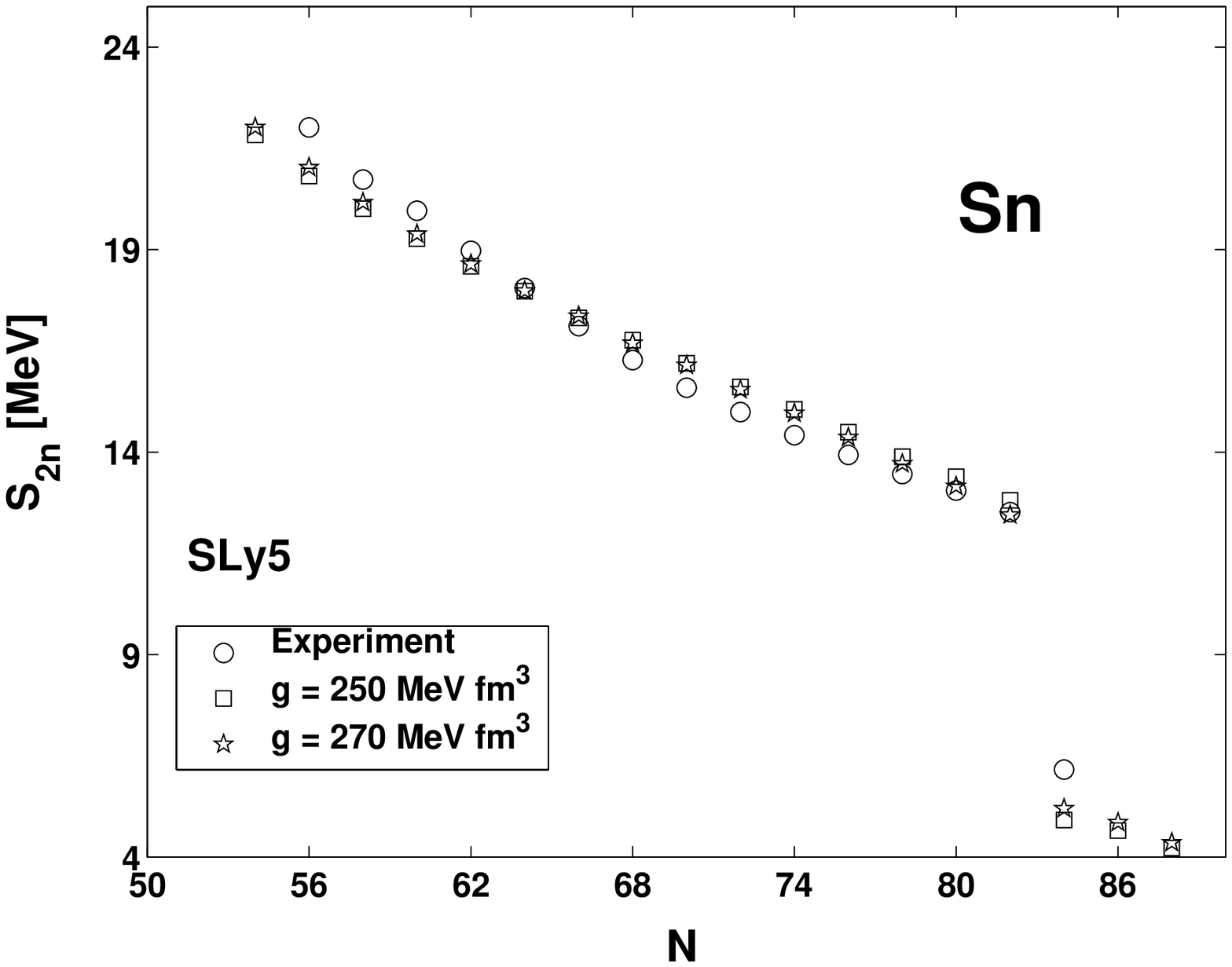,width=4in}
\psfig{figure=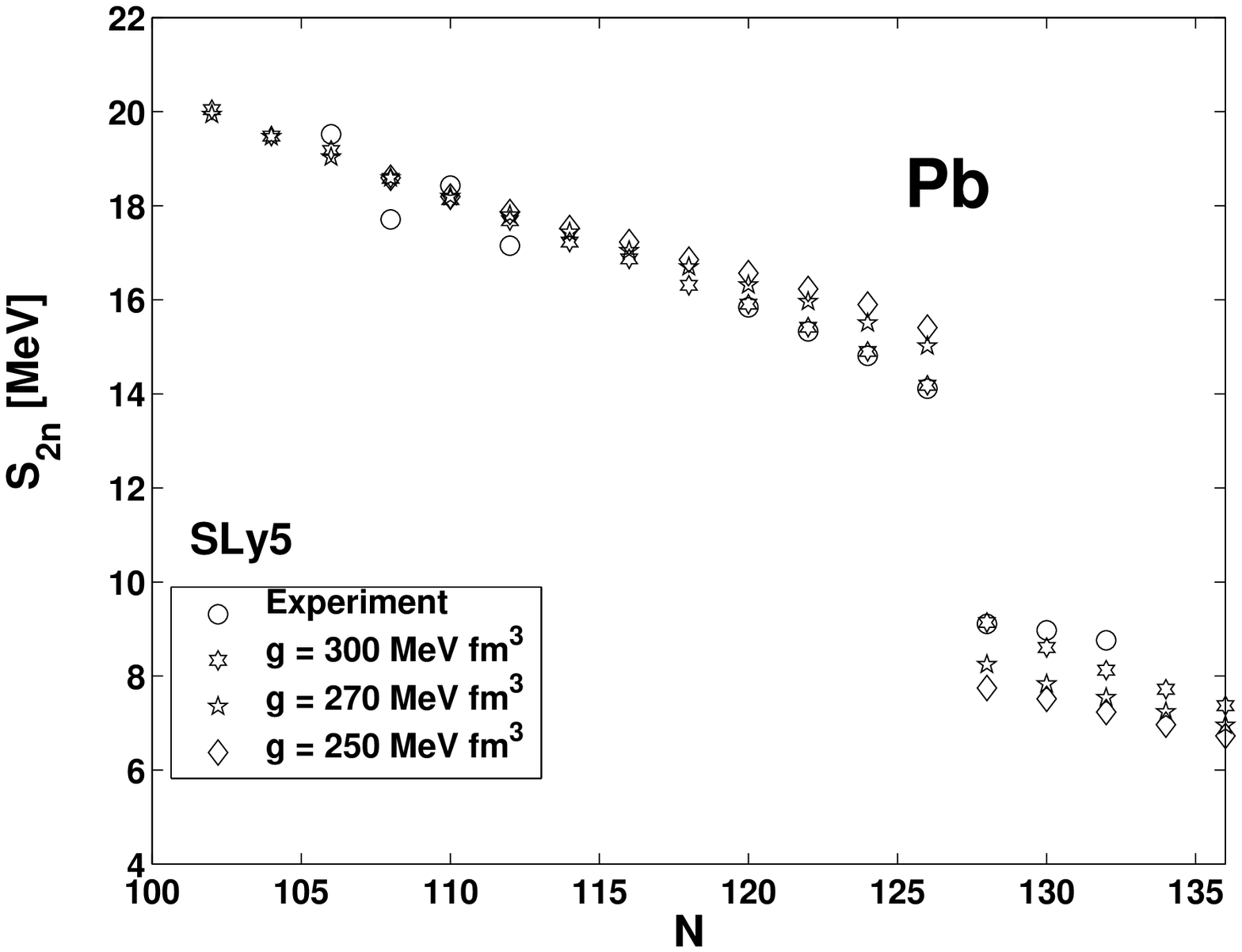,width=4in}
\caption{ The two--neutron separation energies $S_{2n}$ in tin and lead
  isotopes for several values of the bare coupling constant $g$ and the
  corresponding experimental values. }
\label{fig:s2n_pb}
\end{figure}

\section{Conclusions} 

The main issue we have tried to address here is how to proceed in order to
introduce a well defined LDA nuclear energy density functional, which includes
a meaningful description of the pairing properties as well. Until now such a
theoretical framework was missing.  We have shown here that with very little
effort such a framework can be formulated.  The same procedure should be applied to 
relativistic meanfield calculations with pairing correlations \cite{meng}, if one prefers an
energy density functional derived in this manner. However, this is
not the main achievement, as one might hastily conclude. With this new tool in
hands we now understand that the nuclear pairing problem in a way it is much
simpler then previously thought. There is absolutely no physical reason to
perform calculations of the paring gap involving nuclear single--particle
states with energies of the order of 1 GeV, as quite often one might find in
literature. Pairing affects nuclear properties around a very small energy
window around the fermi level, a few MeV's or so, and any physically
meaningful treatment of the pairing phenomena should involve only states in
this energy region.  We have shown that this is indeed the case. Along the way
we have observed that the ensuing formalism is in some ways very ``simple''
and it requires the introduction of one or perhaps at most two effective
coupling constants.  Even if in the final analysis one will find the need for
more parameters, it is very likely that those parameters will be needed in order
to describe the genuine density dependence of the effective pairing
interaction. This situation is to a large extent totally similar to the
Landau theory of fermi liquids and its extension to nuclei by Migdal. As it was
confirmed many times over in subsequent treatments of nuclei using Skyrme
forces, Landau's and Migdal's ideas needed some relatively minor tweaking. 
In the end the
description of nuclear properties within the Skyrme meanfield approximation is
basically the Landau--Migdal theory made selfconsistent and based on a
relatively small number of bare coupling constants. Where the theory needed
completion was the pairing channel, which, in order to make the theory
selfconsistent, required  a regularization procedure.  

With the success of the Hohenberg and Kohn density functional theory, and 
especially the LDA formulation of Kohn and Sham, we come to understand that the
primary ingredient is the energy density functional. The road to construct it
is basically clear. One has to calculate in an {\it ab initio} approach the
properties of infinite homogeneous matter in order to extract the energy density
functional. To this one has to add the calculation of both normal and
superconducting systems, in order to extract the dependence on the anomalous
densities (proton and neutron) as well. There is one more step, the calculation
of the gradient corrections. There is a lot of ambiguity here and even the
condensed matter and chemistry communities, with their significantly larger
human potential, were not able to arrive at a reasonable answer.  Moreover, in
condensed matter systems and chemistry the interaction is well known. One can
however consider a pure phenomenological approach, such as that suggested by
Fayans \cite{sergei}. There is also the question of whether one needs to
consider gradient corrections in the pairing channel. Here the situation is not
clear yet. On one hand, one might argue that since the pairing occurs in a very
narrow energy strip  around the fermi level, the nucleon momenta are always
essentially equal in length to the fermi momentum $p_F$ and thus there should 
be no need to consider any momentum dependent terms in the effective pairing
interaction. (One should not confuse density with momentum dependence.)  On the
other hand, the short range pairing interaction scatters the nucleons in the
Cooper pair at very large angles and the typical momentum transfer is of the
order of the fermi momentum. This seems to indicate that a (specific) momentum dependence
of the pairing interaction perhaps is needed and maybe even required. However,
until one can point to an observable, which is clearly affected by the momentum
dependence of the pairing interaction (we are not aware of any so far), one can
perhaps use the principle of minimum information or maximum entropy  and
proceed with a simple pairing correction to the energy density functional
suggested here. The exceptional value of such an energy density functional is
its universality, that fact that it can and it should be used to describe both
nuclei, irrespective of their $Z/N$ ratio, along with infinite homogeneous and 
inhomogeneous systems.  

There are still a couple of problems, which are not solved yet, and which
might  have a relatively simple and obvious solution. One has to derive
rules for evaluating other observables involving the anomalous density,
e.g. $\alpha$--decay reduced widths. It is not {\it a priori} clear whether
one should simply use $\nu_c({\bf r})$ or $\nu_{reg}({\bf r})$, or maybe even
something else. One should as well extend the regularization procedure to the
linear response theory, which, superficially, seems rather straightforward.

\section*{Acknowledgments}

AB thanks G.F. Bertsch, J. Dobaczewski  and P.--G. Reinhard
for discussions and DOE for financial support
The warm hospitality of N. Takigawa in Sendai and the financial
support of JSPS during the writing of this  manuscript is greatly appreciated.


\section*{References}

\begin{thebibliography}{99}

\bibitem{pines} A. Bohr, B.R. Mottelson and D.Pines, Phys. Rev. {\bf
110}, 936 (1958).

\bibitem{cooper} L.N. Cooper, Phys. Rev. {\bf 104}, 1189 (1956).

\bibitem{bcs} J. Bardeen, L.N. Cooper, J.R. Schrieffer, Phys. Rev. {\bf 108},
  1175 (1957).
 
\bibitem{mohit} M. Randeria, in {\it Bose--Einstein Condensation},
eds. A. Griffin, D.W. Snoke and S. Stringari, Cambridge University
Press (1995), pp 355--392.
 
\bibitem{hk} P. Hohenberg and W. Kohn, Phys. Rev. {\bf 136}, B864
(1964); W. Kohn and L. J. Sham, Phys. Rev. {\bf 140}, A1133 (1965);
R.M. Dreizler and E.K.U. Gross, {\it Density Functional Theory: An
Approach to the
 Quantum Many--Body Problem}, (Springer, Berlin, 1990).
 
\bibitem{negele} J.W. Negele and D. Vautherin, Phys. Rev. C {\bf 5},
1472 (1972); A. Bulgac, C. Lewenkopf and V. Mickrjukov, Phys. Rev. B
{\bf 52}, 16476 (1995).

\bibitem{sergei} S.A. Fayans, JETP Letter. {\bf 68}, 169 (1999); 
 S.A. Fayans {\it et al}, JETP Letter. {\bf 68}, 276 (1999).

 \bibitem{oliveira} L.N. Oliveira, E.K.U. Gross and W. Kohn,
Phys. Rev. Lett. {\bf 60}, 2430 (1988); S. Kurth {\it et al.},
Phys. Rev. Lett. {\bf 83}, 2628 (1999).

\bibitem{ab} A. Bulgac, preprint FT--194--1980, CIP, Bucharest;
nucl-th/9907088.
 
\bibitem{note1} The notations used in Ref. \cite{ab} are slightly
different from those commonly used in literature, which we try to
follow in this work.
 
\bibitem{bruun} G. Bruun, Y. Castin, R. Dum and K. Burnett,
Eur. Phys. J. D {\bf 7}, 433 (1999).

\bibitem{best} A. Bulgac, nucl-th/0108014.
 
\bibitem{elgaroy}  M. Baldo {\it et al.}, Nucl. Phys. {\bf A 515}, 409 (1990);
\O. Elgaroy {\it et al}, Nucl. Phys. {\bf A 604}, 466 (1996).
 
\bibitem{khodel} V.A. Khodel, V.V. Khodel and J.W. Clark, Nucl. Phys.
{\bf A 598}, 390 (1996).

\bibitem{blatt} J.M. Blatt and V.F. Weiskopf, {\it Theoretical Nuclear
Physics}, Wiley, New York (1952), pp. 74--76;
K. Huang, {\it Statistical Mechanics}, John Wiley \& Sons, New York
(1987), pp 230--238. The procedure amounts to the replacement of the
short range potential $V({\bf r})$ according to the simple
prescription $V({\bf r})\psi ({\bf r}) \rightarrow g\delta({\bf r})
\partial_r [r\psi({\bf r})]$, where the coupling constant is
determined by the scattering length $g= 4\pi a \hbar^2/m$. 
The operator $\delta({\bf r}) \partial_r [r\psi({\bf r})]$ is nothing but a
somewhat pompous formulation amounting to exactly the same thing as ``simply
throwing away'' the leading divergent term. One can easily see by applying
it to a simple case, namely $\delta({\bf r}) \partial_r [r(A/r+B +\ldots )]\equiv
\delta({\bf r}) B$. One can include corrections to the leading order term of 
the pseudo--potential by making the replacement $a\rightarrow -f/(1+ikf)$, 
where $f$ is the $s$--wave scattering amplitude. If needed, one can also 
include higher partial waves \cite{yang}.
 
\bibitem{yang} K. Huang and C.N. Yang, Phys. Rev. {\bf 105}, 767
(1957); T.D. Lee and C.N. Yang, Phys. Rev. {\bf 105}, 1119, (1957).
 
\bibitem{gorkov} A.A. Abrikosov, A.P. Gorkov and I.E. Djaloshinski,
{\it Methods of Quantum Field Theory in Statistical Physics}, Dover,
New York (1975) , Ch. 1.5.
 
\bibitem{randeria} M. Randeria, J.--M. Duan, L.--Y. Shieh,
Phys. Rev. B {\bf 41}, 327 (1990); C.A.R. S\'{a} de Melo, M. Randeria,
and J.R. Engelbrecht, Phys. Rev. Lett. {\bf 71}, 3202 (1993).
 
\bibitem{fayans} S. Fayans, JETP Lett. {\bf 70}, 240 (1999);
S.A. Fayans {\it et al.}, Nucl. Phys. {\bf A 676}, 49 (2000).

\bibitem{george} T. Papenbrock and G.F. Bertsch, Phys. Rev. C {\bf
59}, 2052 (1999).
 
\bibitem{hsu} S.D.H. Hsu and J. Hormuzdiar, nucl-th/981101.
 
\bibitem{henning}  H. Esbensen, G.F. Bertsch and K. Hencken,
Phys. Rev. C {\bf 56}, 3054 (1997).

\bibitem{ring} P. Ring and P. Schuck, {\it The Nuclear Many--Body
Problem}, Springer, New York (1980), Ch. 4.

\bibitem{gogny} J. Decharg\'e and D. Gogny, Phys. Rev. C {\bf 21}, 1568 (1980);
J.F. Berger, M. Girod and D. Gogny, Comput. Phys. Comm. {\bf 63}, 365 (1991). 

\bibitem{abyy} A. Bulgac and Y. Yu, nucl-th/0106062. 

\bibitem{gor} L.P. Gorkov and T.K. Melik--Barkhudarov, Sov. Phys. JETP
{\bf 13}, 1018 (1961); H. Heiselberg  {\it et al}, Phys. Rev. Lett. {\bf 85},
2418 (2000); H.--J. Schulze, A. Polls and A. Ramos, Phys. Rev. C {\bf 63},
044310 (2001).

\bibitem{typo} The definition of the propagator used in
Ref. \cite{bruun} differs by a sign from the commonly accepted
one.

\bibitem{jacek} J. Dobaczewski, H. Flocard and J. Treiner, Nucl. Phys.
{\bf A 422}, 103 (1984).

\bibitem{note2} The authors of Ref. \cite{bruun} state that previous
authors, in particular those of Ref. \cite{randeria}, have used
$1/(\veps_i -\mu)$ and not $1/\veps_i$ as a subtraction term. As a matter of
fact other authors have used the  $1/(\veps_i -\mu)$ subtraction term, 
e.g. H.T.C. Stoof {\it et al}, Phys. Rev. Lett. {\bf 76}, 10 (1996). 
One has to note that it is still accurate to use the zero energy scattering
amplitude in the corresponding equations only if the effective range corrections 
are negligible.

\bibitem{sly} E. Chabanat {\it et al.}, Nucl. Phys. {\bf A 627}, 710 (1997); 
Nucl. Phys. {\bf A 635}, 231 (1998); {\it Erratum} Nucl. Phys. {\bf A 643}, 441
(1998).

\bibitem{fpwff} B. Friedman and V.R. Pandharipande
Nucl. Phys. {\bf A 361}, 502 (1981); 
R.B. Wiringa, V. Fiks and A. Fabrocini, Phys. Rev. {\bf 38}, 1010 (1988).

\bibitem{garrido} E. Garrido {\it et al}, Phys. Rev.   C {\bf 60},  064312 (1999);
Phys. Rev. C {\bf 63}, 037304 (2001).

\bibitem{vasya} A. Bulgac and V.R. Shaginyan, Phys. Lett. {\bf B 469}, 1
  (1999);.
   
\bibitem{meng} J. Meng, Phys. Rev. C {\bf 57}, 1229 (1998).

\bibitem{jd} J. Dobaczewski, private communication.

\end{thebibliography}
\end{document}